\documentclass[12pt, draftclsnofoot, onecolumn]{IEEEtran}
\usepackage{geometry}
\ifCLASSINFOpdf

\else

\fi
\usepackage[cmex10]{amsmath}
\usepackage{amssymb}
\usepackage{cite}
\usepackage{graphicx}
\usepackage{array,color}
\usepackage{amsmath}
\usepackage{stfloats}
\usepackage{graphicx}
\usepackage{subfigure}
\usepackage{tabularx}
\usepackage{epsfig,epsf,color,balance,cite}
\usepackage{algorithmic}
\usepackage{algorithm}
\usepackage{bm}
\usepackage{textcomp}

\usepackage{epstopdf}
\usepackage{graphicx}
\usepackage{epstopdf}

\usepackage{bm}
\usepackage{multirow}
\usepackage{amsmath}
\usepackage{amssymb}
\usepackage{graphicx}

\makeatletter

\providecommand{\tabularnewline}{\\}
\floatstyle{ruled}
\newfloat{algorithm}{tbp}{loa}
\providecommand{\algorithmname}{Algorithm}
\floatname{algorithm}{\protect\algorithmname}

\usepackage{cite}
\usepackage{array}
\usepackage{color}\usepackage{stfloats}
\usepackage{subfigure}
\usepackage{tabularx}
\usepackage{epsfig}
\usepackage{epsf}\usepackage{color}\usepackage{balance}\usepackage{cite}\usepackage{algorithmic}
\usepackage{algorithm}
\usepackage{bm}
\usepackage{textcomp}

\usepackage{epstopdf}
\usepackage{epstopdf}

\makeatother

\begin{document}
\title{Robust Transmission Design for Intelligent Reflecting Surface Aided Secure Communication Systems with Imperfect Cascaded CSI}
\author{ Sheng Hong, Cunhua Pan, Hong Ren, Kezhi Wang, Kok Keong Chai, \IEEEmembership{Member, IEEE}, and Arumugam Nallanathan, \IEEEmembership{Fellow, IEEE}
%\thanks{This work was supported by ...}
\thanks{S. Hong is with Information Engineering School of Nanchang University, Nanchang 330031, China. She is also with the School of Electronic Engineering and Computer Science at Queen Mary University of London, London E1 4NS, U.K. (email: shenghong@ncu.edu.cn). C. Pan, H. Ren, K. K. Chai, and A. Nallanathan are with the School of Electronic Engineering and Computer Science at Queen Mary University of London, London E1 4NS, U.K. (e-mail:\{c.pan, h.ren, michael.chai, a.nallanathan\}@qmul.ac.uk). K. Wang is with Department of Computer and Information Sciences, Northumbria University, UK. (email: kezhi.wang@northumbria.ac.uk).
}
}
\maketitle
\vspace{-1.9cm}
\begin{abstract}
In this paper, we investigate the design of robust and secure transmission
in intelligent reflecting surface (IRS) aided wireless communication
systems. In particular, a multi-antenna access point (AP) communicates with a single-antenna
legitimate receiver in the presence of multiple single-antenna eavesdroppers, where the artificial noise (AN) is transmitted to enhance the security performance. Besides, we assume that the cascaded AP-IRS-user channels are imperfect due to the channel estimation error. To minimize
the transmit power, the beamforming vector at the transmitter, the AN covariance
matrix, and the IRS phase shifts are jointly optimized subject
to the outage rate probability constraints under the statistical cascaded channel state information (CSI) error model that usually models the channel estimation error. To handle the resulting non-convex optimization problem,
we first approximate the outage rate probability constraints by using
the Bernstein-type inequality. Then, we develop a suboptimal algorithm
based on alternating optimization, the penalty-based and semidefinite
relaxation methods. Simulation results reveal that the proposed scheme
significantly reduces the transmit power compared to other benchmark
schemes.
\end{abstract}
\section{Introduction}

The intelligent reflecting surface (IRS) has emerged due to the advancement
in metamaterial techniques, which becomes a promising technology in wireless networks \cite{wu2019towards}. The IRS, which is also
named as reconfigurable intelligent surface (RIS), comprises a large
number of passive elements, which can reflect the wireless signal with
adjustable phase shifts \cite{zhao2019survey}\cite{cui2014coding}.
The IRS has the capability of reconfiguring the wireless propagation environment between access point (AP) and users in a favourable way by properly designing
the phase shifts. Thus, the IRS can improve the performance of wireless
networks in various aspects. Since the IRS operates in a passive mode
by reflecting incident signals\cite{ntontin2019reconfigurable}, it
can significantly improve the spectral and energy efficiency of the
wireless networks. Besides, it is very appealing that the IRS is low-cost,
and can be deployed easily on buildings facades, interior walls, room
ceilings, lamp posts and road signs, etc. Therefore, exploiting the IRS device to assist wireless
communication systems has received extensive attentions. The IRS-aided wireless communication systems in the existing literature
include the single-user case \cite{yu2019miso}\cite{yang2020intelligent},
the downlink multiuser case \cite{huang2019reconfigurable}\cite{wu2019intelligent}\cite{guo2019weighted}\cite{nadeem2019large}
\cite{zhou2019intelligent}, multicell scenario \cite{pan2019intelligent2}, wireless power transfer design\cite{pan2019intelligent}, mobile edge computing \cite{bai2019latency}, and cognitive radio system \cite{xu2020resource}.

Recently, the IRS has emerged as a promising technology to enhance
the physical layer security in a wireless communication system.
There are various schemes to improve the physical layer security\cite{shiu2011physical},
such as the cooperative relaying\cite{li2011cooperative}, artificial
noise (AN)\cite{sun2018robust}, and cooperative jamming\cite{dong2009improving}.
By using these schemes, the AP-eavesdroppers links are impaired, and
the information leakage to the eavesdroppers is limited. Combining with these schemes, the IRS can further enhance the physical
layer security. On one hand, deploying the IRS costs much less than deploying the relay since no active radio frequency chain is required in IRS-aided systems. On the other hand, the IRS can be programmed
to configure the radio propagation environment to make the impairment
on the eavesdroppers' channels more effective.

In terms of physical layer security enhancement, the IRS-aided secure communication has received considerable attention from academia \cite{yu2019enabling} \cite{cui2019secure}\cite{shen2019secrecy}\cite{chen2019intelligent}.
In these contributions, the active transmit beamforming and the passive reflecting beamforming of the IRS were jointly designed to improve the achievable
secrecy rate. It was demonstrated that it is preferable to deploy
the IRS near the legitimate receiver. The work in \cite{cui2019secure} assumed that the eavesdropper
has a stronger channel than the legitimate user and both channels
are highly correlated, and it showed that deploying an IRS can enhance
the secrecy rate even in such a challenging scenario. Moreover, the
AN is incorporated with transmit beamforming in IRS-aided wireless
secure communications\cite{guan2019intelligent}\cite{xu2019resource}.
The phase shift matrix at the IRS as well as the beamforming vectors and
AN covariance matrix at the base station (BS) were jointly optimized for maximizing
the secrecy rate. It was unveiled that it is beneficial for secrecy
enhancement with the aid of AN.

We note that all these existing contributions are based on the ideal assumption of perfect channel state information (CSI) of both the
legitimate receiver and the eavesdropper at the transmitters. However,
it is difficult to obtain the perfect CSI in IRS-aided wireless
communication, because the IRS does not employ any radio frequency (RF) chains. In IRS-aided communication systems, it is challenging to estimate the IRS-related
channel of the reflective AP-IRS-user link due to the passive IRS, which thus attracted a lot of research attention. The current channel estimation contributions for the IRS-related channels can be divided into two approaches. The first one is to
estimate AP-IRS channel and IRS-user channel separately \cite{taha2019enabling}.
As shown in \cite{taha2019enabling}, to estimate the two separated
IRS-related channels, active transmit radio RF chains are required to be installed at the IRS. The drawback of this approach is that extra hardware and power consumption is required. The second one is to estimate the cascaded AP-IRS-user
channel, which is defined as the product of the AP-IRS channel and
the IRS-user channel \cite{zhou2019intelligent}\cite{yu2019enabling}\cite{shen2019secrecy}\cite{zhang2019capacity}.
The benefit of this approach is that the cascaded AP-IRS-user channel can be estimated without installing extra hardware and
incurring power cost, and that the estimated cascaded AP-IRS-user channel is sufficient for the transmission beamforming design for the IRS-related links. Thus, more efforts are dedicated to the cascaded channel estimation \cite{unknown}\cite{wang2019channel}\cite{wang2019compressed}\cite{chen2019channel}.
The cascaded channel estimation methods were investigated in a single-user
multiple-input multiple-output (SU-MIMO) system\cite{unknown} and a multi-user multiple-input single-output (MU-MISO) system\cite{wang2019channel}, respectively. Then, by exploiting the channel sparsity of the mmWave channels, the compressive sensing methods were adopted to reduce the pilot overhead in \cite{wang2019compressed} and \cite{chen2019channel} for single-user and multi-user systems, respectively. However, for both approaches, the channel estimation error is inevitable. Therefore, it is imperative to
take the channel estimation error into consideration when designing the IRS-aided communication systems.

There are a paucity of contributions investigating the robust transmission design
in IRS-aided communications. The work of \cite{zhou2019robust} proposed a worst-case robust design algorithm in an MU-MISO
wireless system, where the IRS-user channels were assumed to be imperfect. In addition, a worst-case robust design in IRS-assisted secure
communications was investigated in \cite{yu2019robust},
where the IRS-eavesdropper channels were assumed to be imperfect.
Both these works only considered the IRS-user CSI error based
on the first IRS-related channel estimation approach. Since it is
more practical to consider the cascaded channel uncertainty based on the second IRS-related channel estimation approach, Zhou \emph{et. al}
\cite{zhou2020framework} firstly proposed a robust transmission
framework with both the bounded and the statistical cascaded CSI
errors in an MU-MISO wireless system.

However, to the best of our knowledge, the robust transmission design with cascaded channel error in secure IRS-aided communication systems has not been studied yet. Moreover, the probabilistic CSI error model has not been studied in secure IRS-aided communication systems. To fill this gap, this paper investigates the outage constrained
robust secure transmission for IRS-aided secure wireless communication systems, where the statistical model of the cascaded channel error is considered. Specifically, we consider a scenario that a multi-antenna AP serves
a single-antenna legitimate user in the presence of several single-antenna
eavesdroppers. Moreover, the AN is assumed to be injected to degrade the reception quality
of eavesdroppers. In this scenario, we consider an outage-constrained
robust design aiming for minimizing the transmit power by considering the
imperfect CSI of the  eavesdroppers' channels with the outage constraints
of maximum information leakage to eavesdroppers and the constraint
of minimum information transmission to legitimate users. An outage-constrained
power minimization (OC-PM) problem is formulated to jointly optimize the beamforming
vector, the AN covariance matrix, and the phase shifts matrix of IRS
while satisfying quality of service (QoS) requirements. Since the optimization
variables are highly coupled, an alternating optimization (AO) method
is proposed to solve it.

The main contributions of this paper can be summarized as follows:
\begin{enumerate}
\item To the best of our knowledge, this is the first work to study the
outage constrained robust secure transmission for IRS-aided wireless communications. In contrast to \cite{zhou2019robust} and \cite{yu2019robust} that only considered the bounded CSI error model in secure IRS-aided
wireless communications, we first consider the statisitical CSI error model. In addition, the imperfect cascaded channels of
AP-eavesdropper are investigated in secure IRS-aided communications
for the first time, which is more practical compared to the existing literature
considering the imperfect IRS-user channels.
\item For the outage constrained robust secure transmission design, we formulate
an OC-PM problem to optimize
the beamforming vector, the AN covariance matrix, and the phase shifts
matrix of IRS. To solve it, we first transform the probabilistic constraints into
the safe and tractable forms by exploiting Bernstein-type inequality (BTI) \cite{wang2014outage}.
Then, the AO strategy is utilized to transform the original problems into two subproblems,
where the designs of beamforming vector, AN covariance matrix, and
phase shifts of IRS are handled by semidefinite
relaxation (SDR) based methods. For the nonconvex
unit modulus constraints of IRS phase shifts, an equivalent rank-one
constraint is incorporated, which is further transformed by the first-order
Taylor approximation and added into the objective function as a penalty.
\item Simulation results demonstrate that the robust design of beamforming
vector and AN covariance matrix in a secure IRS-aided communication system can reduce the transmit power under the fixed secrecy rate.
In comparison to various benchmark methods, the effectiveness of the
proposed AO algorithm is verified. Moreover, it is revealed that by deploying the IRS and optimizing the IRS phase shifts, the reliable communication
can be guaranteed for the legitimate user, while the information leakage
to eavesdropper can be limited. The physical layer security can be significantly enhanced
by the robust design in the secure IRS-aided wireless communications.
\end{enumerate}

\emph{Notations}: Throughout this paper, boldface lower case, boldface upper case and regular letters are used to denote vectors, matrices, and scalars, respectively. ${\rm{Tr}}\left( {\bf{X}} \right)$ and $\left| {\bf{X}} \right|$ denote the trace and determinant of ${\bf{X}}$, respectively. ${{\mathbb{ C}}^{M \times N}}$ denotes the space of $M \times N$ complex matrices. ${\rm{Re}}\{\cdot\}$ denotes the real part of a complex value. ${\rm{diag}}(\cdot)$ is the operator for diagonalization. $\mathbf{1}_{M}$ presents a column vector with $M$ ones. ${\cal C}{\cal N}({\bm{\mu}},{\bf{Z}})$ represents a circularly symmetric complex gaussian (CSCG) random vector with mean ${\bm{\mu}}$ and covariance matrix ${\bf{Z}}$. ${\left( \cdot \right)^{\rm{T}}}$, ${\left(  \cdot \right)^{\rm{H}}}$ and ${\left( \cdot \right)^{\rm{\ast}}}$ denote the transpose, Hermitian and conjugate operators, respectively.

\section{System model}

In this section, we present the transmission model and CSI error model
in a secure IRS-aided communication system as follows.

\subsection{Signal Transmission Model}

An IRS-assisted communication system is considered, where an AP equipped
with $N_{t}$ antennas, called Alice, intends to send confidential
information to a single-antenna legitimate receiver, called Bob, in
the presence of $K$ single-antenna eavesdroppers, called Eves.
\begin{figure}[h!]
\centering
\includegraphics[width=3.5in]{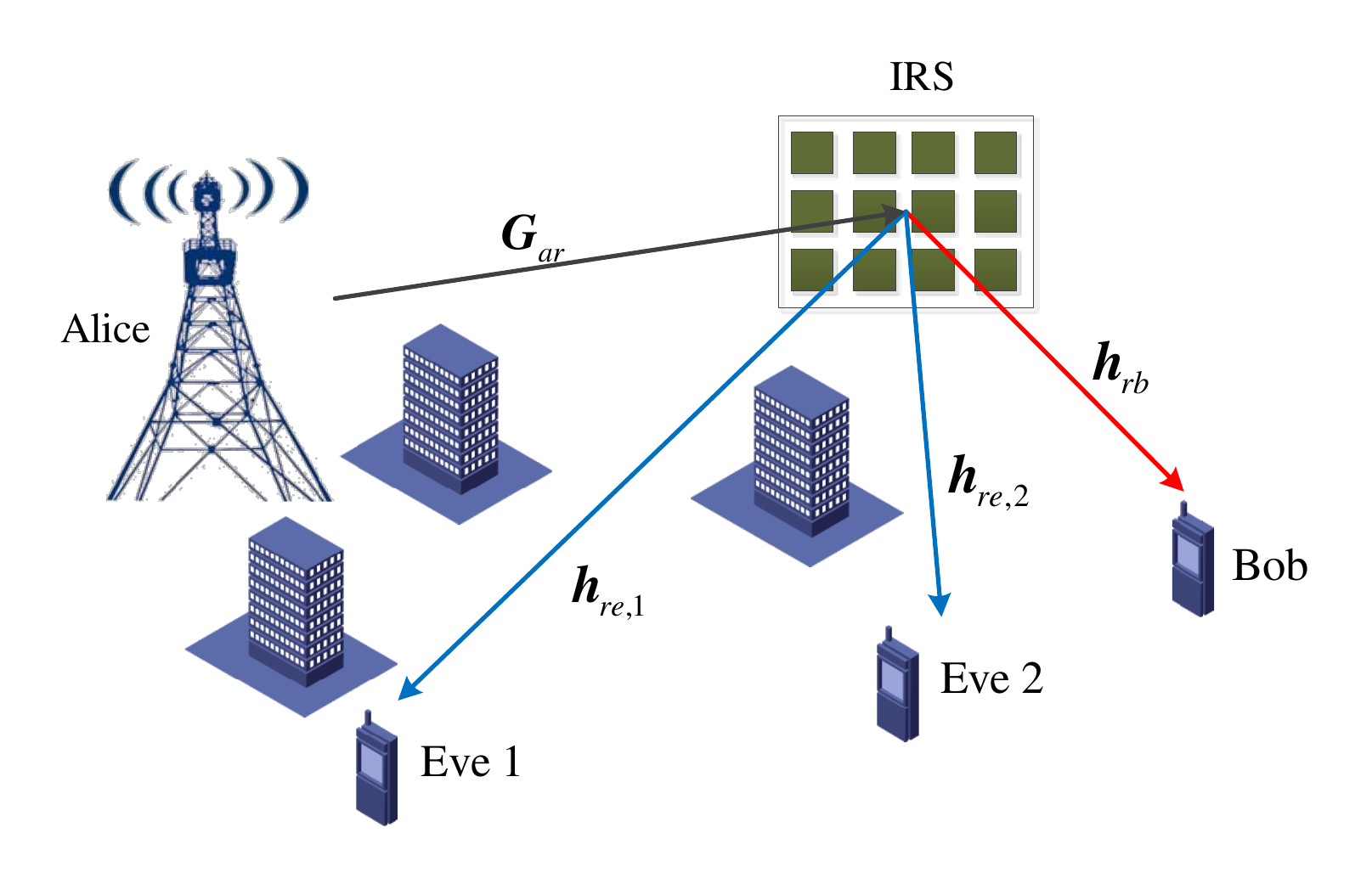}
\caption{An IRS-assisted secure communication system.}\vspace{-0.8cm}
\label{figsystemmodel}
\end{figure}

As shown in Fig. \ref{figsystemmodel}, the signal transmitted from Alice
is reflected by the IRS, which comprises $M$ reflecting elements. The
direct links of Alice-Bob and Alice-Eve are blocked by obstacles such
as buildings. The Bob and Eves can only receive the signals reflected
by the IRS. The received signals at Bob and Eves are respectively modeled as
\begin{equation}
y_{b}(t)=\hat{\mathbf{h}}_{b}^{H}\mathbf{x}(t)+n_{b}(t)=(\mathbf{\mathbf{h}}_{rb}^{H}\mathbf{\Phi}\mathbf{G}_{ar})\mathbf{x}(t)+n_{b}(t),
\end{equation}
\begin{equation}
\mathbf{\textrm{y}}_{e,k}(t)=\mathbf{\hat{h}}_{e,k}^{H}\mathbf{x}(t)+\mathbf{\textrm{n}}_{e,k}(t)=(\mathbf{h}_{re,k}^{H}\mathbf{\Phi}\mathbf{G}_{ar})\mathbf{x}(t)+n_{e,k}(t),
\end{equation}
where $\hat{\mathbf{h}}_{b}=\mathbf{G}_{ar}^{H}\mathbf{\mathbf{\Phi}}^{H}\mathbf{h}_{rb}$, $\hat{\mathbf{h}}_{b}\in\mathbb{C}^{N_{t}\times1}$ is defined as the equivalent channel spanning from Alice to Bob. The channels of the IRS-Bob link and the Alice-IRS link are modeled by $\mathbf{h}_{rb}\in\mathbb{\mathbb{C}}^{M\times1}$ and $\mathbf{\mathbf{G}}_{ar}\in\mathbb{\mathbb{C}}^{M\times N_{t}}$, respectively. $\mathbf{\hat{h}}_{e,k}=\mathbf{G}_{ar}^{H}\mathbf{\mathbf{\Phi}}^{H}\mathbf{h}_{re,k}$, $\mathbf{\hat{h}}_{e,k}\in\mathbb{C}^{N_{t}\times1}$ is defined as the equivalent channel spanning from Alice to the
$k$th Eve. The channel of the IRS-Eve link is modeled by $\mathbf{h}_{re,k}\in\mathbb{\mathbb{C}}^{M\times1}$. $n_{b}(t)$ is the additive white Gaussian noise (AWGN) at Bob with
zero mean and noise variance $\sigma_{b}^{2}$, i.e., $n_{b}\sim\mathcal{CN}(0,\sigma_{b}^{2})$.
$n_{e,k}(t)$ is the AWGN at the $k$th Eve with zero mean and noise covariance
matrix $\sigma_{e,k}^{2}$, i.e., $n_{e,k}\sim\mathcal{CN}(0,\sigma_{e,k}^{2})$.
The phase shift coefficients of the IRS are collected in a diagonal matrix
defined by $\mathbf{\Phi}={\rm{diag}}\{\phi_{1},\cdots,\phi_{m},\cdots,\phi_{M}\}$,
where $\phi_{m}=$$e^{j\theta_{m}}$, and $\theta_{m}\in[0,2\pi]$
denotes the phase shift of the $m$-th reflecting element.

The transmit signal vector is
\begin{equation}
\mathbf{x}(t)=\mathbf{s}(t)+\mathbf{z}(t)=\mathbf{w}s(t)+\mathbf{z}(t),
\end{equation}
where $s(t)$ is a data stream carrying the confidential information
intended for Bob, $\mathbf{z}(t)$ is the noise vector artificially
created by Alice to interfere Eves, i.e., the so-called AN, and $\mathbf{w}$
is the beamforming vector. The confidential signal vector $\mathbf{s}(t)$
is assumed to follow a complex Gaussian distribution of $\mathcal{CN}(\mathbf{0},\mathbf{W})$,
where $\mathbf{W=}\mathbf{w}\mathbf{w}^{H}$ is the transmit covariance matrix to be designed. For the AN, we assume $\mathbf{z}(t)\sim\mathcal{CN}(\mathbf{0},\mathbf{Z})$,
where $\mathbf{Z}$ is the AN covariance matrix to be designed.

The equivalent channel $\hat{\mathbf{h}}_{b}$ can be rewritten as
\begin{equation}
\hat{\mathbf{h}}_{b}^{H}=\bm{\phi}^{T}\textrm{diag(\ensuremath{\mathbf{\mathbf{h}}_{rb}^{H}})}\mathbf{G}_{ar}\triangleq\bm{\phi}^{T}\mathbf{G}_{cb},
\end{equation}
where $\bm{\phi}=[\phi_{1},\cdots,\phi_{m},\cdots,\phi_{M}]^{T}$
and $\mathbf{G}_{cb}=\textrm{diag(\ensuremath{\mathbf{\mathbf{h}}_{rb}^{H}})}\mathbf{G}_{ar}$. $\mathbf{G}_{cb}\in\mathbb{C}^{M\times N_{t}}$ is defined as the cascaded channel from Alice to Bob via the IRS. Similarly, the
equivalent channel $\mathbf{\hat{h}}_{e,k}$ can be expressed in another
way as
\begin{equation}
\hat{\mathbf{h}}_{e,k}^{H}=\bm{\phi}^{T}\textrm{diag(\ensuremath{\mathbf{\mathbf{h}}_{re,k}^{H}})}\mathbf{G}_{ar}\triangleq\bm{\phi}^{T}\mathbf{G}_{ce,k},
\end{equation}
where $\mathbf{G}_{ce,k}=\textrm{diag(\ensuremath{\mathbf{\mathbf{h}}_{re,k}^{H}})}\mathbf{G}_{ar}$,
$k=1,2,\cdots,K$. $\mathbf{G}_{ce,k}\in\mathbb{C}^{M\times N_{t}}$ denotes the cascaded channel from Alice to the $k$th
Eve via the IRS.

\subsection{Statistical CSI Error Model}

In IRS-aided communications, the cascaded AP-IRS-user channels
at the transmitter is challenging to obtain due to the passive features
of the IRS. Hence, we consider the CSI error in the cascaded AP-IRS-user
channels. Generally, there are two channel error models: the bounded
CSI error model and the statistical CSI error model. In this paper,
we focus on the statistical CSI error of the cascaded channel for
Eves since the bounded CSI error is more conservative. Furthermore, the statistical CSI error model is more closely related to the channel estimation error, while the bounded CSI error usually models the quantization error.

We assume the cascaded channel $\mathbf{G}_{ce,k}$ from Alice
to the $k$th Eve is imperfect, which can be represented as
\begin{equation}
\mathbf{G}_{ce,k}=\mathbf{\bar{\mathbf{G}}}_{ce,k}+\triangle\mathbf{G}_{ce,k},\mathbf{g}_{ce,k}\triangleq\textrm{vec}(\triangle\mathbf{G}_{ce,k})\sim\mathcal{CN}(\mathbf{0},\mathbf{\Sigma}_{e,k}),
\end{equation}
where the CSI error vector $\mathbf{g}_{ce,k}\triangleq\textrm{vec}(\triangle\mathbf{G}_{ce,k})$
is assumed to follow the CSCG distribution, and $\mathbf{\Sigma}_{e,k}\in\mathbb{\mathbb{C}}^{MN_{t}\times MN_{t}}$
is the semidefinite error covariance matrix.
\section{Problem Formulation }
In this section, we first derive the secrecy rate expression and then formulate the outage-constrained power minimization problem.
\subsection{Secrecy Rate}
The achievable data rate (bit/s/Hz) of Bob is given by
\begin{align}
C_{b}(\mathbf{W},\mathbf{Z},{\bf \Phi}) & =\log\left(1+\frac{\hat{\mathbf{h}}_{b}^{H}\mathbf{W}\hat{\mathbf{h}}_{b}}{\sigma_{b}^{2}+\hat{\mathbf{h}}_{b}^{H}\mathbf{Z}\hat{\mathbf{h}}_{b}}\right) \nonumber \\
 & =\log\left(1+\frac{(\bm{\phi}^{T}\mathbf{G}_{cb})\mathbf{W}(\bm{\phi}^{T}\mathbf{G}_{cb})^{H}}{\sigma_{b}^{2}+(\bm{\phi}^{T}\mathbf{G}_{cb})\mathbf{Z}(\bm{\phi}^{T}\mathbf{G}_{cb})^{H}}\right),\label{eq:Cb}
\end{align}
where the cascaded channel $\mathbf{G}_{cb}$ from Alice
to Bob is utilized. The achievable data rate (bit/s/Hz) of the $k$th
Eve is given by
\begin{align}
C_{e,k}(\mathbf{W},\mathbf{Z},{\bf \Phi}) & =\log\left(1+\frac{\hat{\mathbf{h}}_{e,k}^{H}\mathbf{W}\hat{\mathbf{h}}_{e,k}}{\sigma_{e,k}^{2}+\hat{\mathbf{h}}_{e,k}^{H}\mathbf{Z}\hat{\mathbf{h}}_{e,k}}\right) \nonumber \\
&=\log\left(1+\frac{(\bm{\phi}^{T}\mathbf{G}_{ce,k})\mathbf{W}(\bm{\phi}^{T}\mathbf{G}_{ce,k})^{H}}{\sigma_{e,k}^{2}+(\bm{\phi}^{T}\mathbf{G}_{ce,k})\mathbf{Z}(\bm{\phi}^{T}\mathbf{G}_{ce,k})^{H}}\right),\label{eq:Cek}
\end{align}
where the cascaded channel $\mathbf{G}_{ce,k}$ from
Alice to the $k$th Eve is applied.

Then the achievable secrecy rate is
\begin{alignat}{1}
R_{s}(\mathbf{W},\mathbf{Z},{\bf \Phi}) & =\underset{k=1,2,\cdots,K}{\min}\left\{ C_{b}(\mathbf{W},\mathbf{Z},{\bf \Phi})-C_{e,k}(\mathbf{W},\mathbf{Z},{\bf \Phi})\right\}.
\end{alignat}
\subsection{Power Minimization}
In this paper, we aim to jointly optimize the transmit beamforming vector $\mathbf{w}$ (or the transmit beamforming
matrix $\mathbf{W}=\mathbf{w}\mathbf{w}^{H}$), the covariance
matrix $\mathbf{Z}$ of AN and the IRS phase shifts ${\bf \Phi}$ to minimize
the transmit power subject to the Bob's data rate requirement and
leaked data rate outage for Eves. The OC-PM problem can be formulated as
\begin{subequations}\label{OS_powermini}
\begin{alignat}{1}
\underset{\mathbf{W},\mathbf{Z},{\bf \Phi}}{\min} & {\rm Tr(}{\bf W}{\rm +}{\bf Z}{\rm )}\label{eq:OS_powermin_obj}\\
\textrm{s.t.} \quad & C_{b}(\mathbf{W},\mathbf{Z},{\bf \Phi})\geq\log\gamma,\label{eq:bobrate}\\
\textrm{} & \textrm{Pr}_{\mathbf{g}_{ce,k}}\left\{ C_{e,k}(\mathbf{W},\mathbf{Z},{\bf \Phi})\leq\log\beta\right\} \geq1-\rho_{k},k=1,2,\cdots,K,\label{eq:OS_everate}\\
 & {\bf Z}\succeq0,\label{eq:OS_powermin_Z}\\
 & {\bf W}\succeq0,\label{eq:OS_powermin_W}\\
 & \textrm{rank}({\bf W})=1,\label{eq:OS_powermin_Wrank1}\\
 &\left|\phi_{m}\right|=1,m=1,\cdots,M, \label{eq:OS_powermin_unimodulus}
\end{alignat}
\end{subequations}
where $\gamma$, $\beta$, and ${\rho_k, \forall k}$ are constant values, and \eqref{eq:OS_everate} can be seen as the constraints of per-Eve secrecy outage
probability.
\section{Problem solution}
Due to the coupling relation of various optimization variables and the complicated probabilistic constraints,
the proposed OC-PM problem is nonconvex and challenging to solve. To tackle it, we first transform the probabilistic constraints into
the safe and tractable forms. Then, the AO strategy is utilized to decouple the parameters and transform the original problems into two subproblems.
\subsection{Problem Reformulation}
We first reformulate the constraint in \eqref{eq:bobrate}. Specifically, it can be simplified by removing the log function as
\begin{subequations}
\begin{alignat}{1}
 & \log\left(1+\frac{\hat{\mathbf{h}}_{b}^{H}\mathbf{W}\hat{\mathbf{h}}_{b}}{\sigma_{b}^{2}+\hat{\mathbf{h}}_{b}^{H}\mathbf{Z}\hat{\mathbf{h}}_{b}}\right)\geq\log\gamma\\
\Leftrightarrow & \textrm{Tr}([\mathbf{W}-(\gamma-1)\mathbf{Z}]\hat{\mathbf{h}}_{b}\hat{\mathbf{h}}_{b}^{H})\geq(\gamma-1)\sigma_{b}^{2},\\
\Leftrightarrow & \textrm{Tr}([\mathbf{W}-(\gamma-1)\mathbf{Z}](\mathbf{G}_{cb}^{H}\bm{\phi}^{*})(\bm{\phi}^{T}\mathbf{G}_{cb}))\geq(\gamma-1)\sigma_{b}^{2},\\
\Leftrightarrow & \textrm{Tr}(\mathbf{G}_{cb}[\mathbf{W}-(\gamma-1)\mathbf{Z}]\mathbf{G}_{cb}^{H}\mathbf{E})\geq(\gamma-1)\sigma_{b}^{2}, \label{eq:bobdatarate_c}
\end{alignat}
\end{subequations}
where $\mathbf{E}\triangleq\bm{\phi}^{*}\bm{\phi}^{T}$. Then, we reformulate
the per-Eve outage probability constraint \eqref{eq:OS_everate}, which is not tractable to handle due to the log function and the probability requirement. We first eliminate
the log function, then transform the probability constraint into the
deterministic constraint. There are two steps to reformulate the chance
constraint \eqref{eq:OS_everate} as follows.

Step 1): Firstly, the data rate leakage expression in \eqref{eq:OS_everate}
can be simplifed as
\begin{subequations}\label{everateinequality}
\begin{flalign}
 & \log\left(1+\frac{\hat{\mathbf{h}}_{e,k}^{H}\mathbf{W}\hat{\mathbf{h}}_{e,k}}{\sigma_{e,k}^{2}+\hat{\mathbf{h}}_{e,k}^{H}\mathbf{Z}\hat{\mathbf{h}}_{e,k}}\right)\leq\log\beta\label{eq:everateinequality_a}\\
\Leftrightarrow & \hat{\mathbf{h}}_{e,k}^{H}[\mathbf{W}-(\beta-1)\mathbf{Z}]\hat{\mathbf{h}}_{e,k}\leq(\beta-1)\sigma_{e,k}^{2}.\label{eq:everateinequality_b}
\end{flalign}
\end{subequations}

By substituting $\mathbf{\hat{h}}_{e,k}^{H}=\bm{\phi}^{T}(\mathbf{\bar{\mathbf{G}}}_{ce,k}+\triangle\mathbf{G}_{ce,k})$
into \eqref{eq:everateinequality_b} and defining $\mathbf{\Xi}_{e}\triangleq(\beta-1)\mathbf{Z}-\mathbf{W}$,
we have
\begin{subequations}
\begin{flalign}
\eqref{eq:everateinequality_a}\Leftrightarrow & [\bm{\phi}^{T}(\mathbf{\bar{\mathbf{G}}}_{ce,k}+\triangle\mathbf{G}_{ce,k})]\mathbf{\Xi}_{e}[\bm{\phi}^{T}(\mathbf{\bar{\mathbf{G}}}_{ce,k}+\triangle\mathbf{G}_{ce,k})]^{H}+(\beta-1)\sigma_{e,k}^{2}\geq0,\label{eq:evedatarate_a}\\
\Leftrightarrow & \underset{f_{1}}{\underbrace{\bm{\phi}^{T}\triangle\mathbf{G}_{ce,k}\mathbf{\Xi}_{e}\triangle\mathbf{G}_{ce,k}^{H}\bm{\phi}^{*}}}+2\textrm{Re}\{\underset{f_{2}}{\underbrace{\bm{\phi}^{T}\triangle\mathbf{G}_{ce,k}\mathbf{\Xi}_{e}\mathbf{\bar{\mathbf{G}}}_{ce,k}^{H}\bm{\phi}^{*}}}\} \nonumber \\
 & +\underset{c_{k}}{\underbrace{(\bm{\phi}^{T}\mathbf{\bar{\mathbf{G}}}_{ce,k})\mathbf{\Xi}_{e}(\bm{\phi}^{T}\mathbf{\bar{\mathbf{G}}}_{ce,k})^{H}+(\beta-1)\sigma_{e,k}^{2}}}\geq0.\label{eq:evedatarate_b}
\end{flalign}
\end{subequations}
Let us equivalently represent the CSI error $\mathbf{g}_{ce,k}=\textrm{vec}(\triangle\mathbf{G}_{ce,k})$ as
$\mathbf{g}_{ce,k}=\mathbf{\Sigma}_{e,k}^{1/2}\mathbf{v}_{ce,k}$,
where $\mathbf{v}_{ce,k}\sim\mathcal{CN}(\mathbf{0},\mathbf{I}_{MN_{t}})$, $\mathbf{\Sigma}_{e,k}=\mathbf{\Sigma}_{e,k}^{1/2}\mathbf{\Sigma}_{e,k}^{1/2}$ and $(\mathbf{\Sigma}_{e,k}^{1/2})^{H}=\mathbf{\Sigma}_{e,k}^{1/2}$.
The expression $f_{1}$ in \eqref{eq:evedatarate_b} can be reformulated as
\begin{alignat}{1}
f_{1} & =\textrm{Tr}(\triangle\mathbf{G}_{ce,k}\mathbf{\Xi}_{e}\triangle\mathbf{G}_{ce,k}^{H}\bm{\phi}^{*}\bm{\phi}^{T})\nonumber \\
 & =\textrm{Tr}(\triangle\mathbf{G}_{ce,k}^{H}\mathbf{E}\triangle\mathbf{G}_{ce,k}\mathbf{\Xi}_{e})\nonumber \\
 & \stackrel{(a)}{=}\textrm{vec}^{H}(\triangle\mathbf{G}_{ce,k})(\mathbf{\Xi}_{e}^{T}\otimes\mathbf{E})\textrm{vec}(\triangle\mathbf{G}_{ce,k})\nonumber \\
 & =\mathbf{g}_{ce,k}^{H}(\mathbf{\Xi}_{e}^{T}\otimes\mathbf{E})\mathbf{g}_{ce,k}\nonumber \\
 & =\mathbf{v}_{ce,k}^{H}\mathbf{\Sigma}_{e,k}^{1/2}(\mathbf{\Xi}_{e}^{T}\otimes\mathbf{E})\mathbf{\Sigma}_{e,k}^{1/2}\mathbf{v}_{ce,k}\nonumber \\
 & \triangleq\mathbf{v}_{ce,k}^{H}\mathbf{A}_{k}\mathbf{v}_{ce,k},\label{eq:f1}
\end{alignat}
where $\mathbf{A}_{k}\triangleq\mathbf{\Sigma}_{e,k}^{1/2}(\mathbf{\Xi}_{e}^{T}\otimes\mathbf{E})\mathbf{\Sigma}_{e,k}^{1/2}$,
and the $(a)$ is obtained by invoking the identity $\textrm{Tr}(\mathbf{A}^{H}\mathbf{BCD})=\textrm{vec}^{H}(\mathbf{A})(\mathbf{D}^{T}\otimes\mathbf{B})\textrm{vec}(\mathbf{C})$.
The expression $f_{2}$ in \eqref{eq:evedatarate_b} can be reformulated as
\begin{alignat}{1}
f_{2} & =\textrm{Tr}(\triangle\mathbf{G}_{ce,k}\mathbf{\Xi}_{e}(\mathbf{\bar{\mathbf{G}}}_{ce,k}^{H}\bm{\phi}^{*})\bm{\phi}^{T})\nonumber \\
 & =\textrm{Tr}(\triangle\mathbf{G}_{ce,k}\mathbf{\Xi}_{e}(\mathbf{\bar{\mathbf{G}}}_{ce,k}^{H}\mathbf{E}))\nonumber \\
 & \overset{(b)}{=}\textrm{vec}^{H}(\mathbf{E}\mathbf{\bar{\mathbf{G}}}_{ce,k})(\mathbf{\Xi}_{e}^{T}\otimes\mathbf{I}_{M})\textrm{vec}(\triangle\mathbf{G}_{ce,k})\nonumber \\
 & =\textrm{vec}^{H}(\mathbf{E}\mathbf{\bar{\mathbf{G}}}_{ce,k})(\mathbf{\Xi}_{e}^{T}\otimes\mathbf{I}_{M})\mathbf{g}_{ce,k}\nonumber \\
 & =\textrm{vec}^{H}(\mathbf{E}\mathbf{\bar{\mathbf{G}}}_{ce,k})(\mathbf{\Xi}_{e}^{T}\otimes\mathbf{I}_{M})\mathbf{\Sigma}_{e,k}^{1/2}\mathbf{v}_{ce,k}\nonumber \\
 & \triangleq\mathbf{u}_{ce,k}^{H}\mathbf{v}_{ce,k},\label{eq:f2}
\end{alignat}
where $\mathbf{u}_{ce,k}^{H}\triangleq\textrm{vec}^{H}(\mathbf{E}\mathbf{\bar{\mathbf{G}}}_{ce,k})(\mathbf{\Xi}_{e}^{T}\otimes\mathbf{I}_{M})\mathbf{\Sigma}_{e,k}^{1/2}$,
and the $(b)$ is obtained by invoking the identity $\textrm{Tr}(\mathbf{A}\mathbf{B}\mathbf{C}^{H})=\textrm{vec}^{H}(\mathbf{C})(\mathbf{B}^{T}\otimes\mathbf{I})\textrm{vec}(\mathbf{A})$.
The expression $c_{k}$ in \eqref{eq:evedatarate_b} can be reformulated as
\begin{equation}
c_{k}=\textrm{Tr}[\mathbf{\bar{\mathbf{G}}}_{ce,k}\mathbf{\Xi}_{e}\mathbf{\bar{\mathbf{G}}}_{ce,k}^{H}\mathbf{E}]+(\beta-1)\sigma_{e,k}^{2}.\label{eq:ck}
\end{equation}

By substituting \eqref{eq:f1}, \eqref{eq:f2} and \eqref{eq:ck}
into \eqref{eq:evedatarate_b}, we have
\begin{alignat}{1}
\eqref{eq:everateinequality_a} & \Leftrightarrow\mathbf{v}_{ce,k}^{H}\mathbf{A}_{k}\mathbf{v}_{ce,k}+2\textrm{Re}\{\mathbf{u}_{ce,k}^{H}\mathbf{v}_{ce,k}\}+c_{k}\geq0.
\end{alignat}

The leakage data rate constraint in \eqref{eq:OS_everate} becomes
\begin{alignat}{2}
\eqref{eq:OS_everate} & \Leftrightarrow & \textrm{Pr}_{\mathbf{v}_{ce,k}}\left\{ \mathbf{v}_{ce,k}^{H}\mathbf{A}_{k}\mathbf{v}_{ce,k}+2\textrm{Re}\{\mathbf{u}_{ce,k}^{H}\mathbf{v}_{ce,k}\}+c_{k}\geq0\right\} \geq1-\rho_{k},k=1,2,\cdots,K.\label{eq:Gussianvectoreql}
\end{alignat}

Step 2): the chance constraint is transformed into a deterministic
constraint by using the BTI, which provides
a safe approximation for \eqref{eq:Gussianvectoreql}.

The equivalence in \eqref{eq:Gussianvectoreql} implies that the outage
probability in \eqref{eq:OS_everate} can be characterized by the
quadratic inequality with respect to (w.r.t.) the Gaussian random
vector $\mathbf{v}_{ce,k}$. Generally, the Bernstein-type inequality
is utilized to approximate a chance constraint safely, which is given
in the following Lemma.

\newtheorem{lemma}{Lemma}

\begin{lemma}\label{Lemma_BIT}

(Bernstein-Type Inequality) For any $(\mathbf{A},\mathbf{u},c)\in\mathbb{H}^{n}\times\mathbb{C}^{n}\times\mathbb{R}$,$\mathbf{v}\sim\mathcal{CN}(\mathbf{0},\mathbf{I}_{n})$
and $\rho\in\text{(0,1]}$, the following implication holds:
\begin{flalign*}
 & \textrm{Pr}_{\mathbf{v}}\left\{ \mathbf{v}^{H}\mathbf{A}\mathbf{v}+2\textrm{Re}\{\mathbf{u}^{H}\mathbf{v}\}+c\geq0\right\} \geq1-\rho\\
\Longleftrightarrow & \textrm{Tr}(\mathbf{A})-\sqrt{2\ln(1/\rho)}\sqrt{\left\Vert \mathbf{A}\right\Vert _{F}^{2}+2\left\Vert \mathbf{u}\right\Vert ^{2}}+\ln(\rho)\cdot\lambda^{+}(-\mathbf{A})+c\geq0,\\
\Longleftrightarrow & \begin{cases}
\textrm{Tr}(\mathbf{A})-\sqrt{-2\ln(\rho)}\cdot x+\ln(\rho)\cdot y+c\geq0,\\
\left\Vert \left[\begin{array}{c}
\textrm{vec}(\mathbf{A}),\\
\sqrt{2}\mathbf{u}
\end{array}\right]\right\Vert _{2}\leq x,\\
y\mathbf{I}_{n}+\mathbf{A}\succeq\mathbf{0},y\geq0,
\end{cases}
\end{flalign*}
where $\lambda^{+}(\mathbf{A})=\max(\lambda_{\max}(\mathbf{A}),0)$.
$x$ and $y$ are slack variables.

\end{lemma}

By invoking Lemma \ref{Lemma_BIT} and introducing the slack variables
$\mathbf{x}=[x_{1},x_{2},\cdots,x_{K}]^{T}$ and $\mathbf{y}=[y_{1},y_{2},\cdots,y_{K}]^{T}$,
we arrive at the desired safe approximation of OC-PM, which is shown
as
\begin{subequations}\label{BITOCPM}
\begin{alignat}{1}
\underset{\mathbf{W},\mathbf{Z},{\rm \bm{\phi}},\mathbf{A},\mathbf{x},\mathbf{y}}{\min} & {\rm Tr(}{\bf W}{\rm +}{\bf Z}{\rm )}\label{eq:BITOCPM_a}\\
\textrm{s.t.}\quad & \textrm{Tr}(\mathbf{A}_{k})-\sqrt{-2\ln(\rho_{k})}\cdot x_{k}+\ln(\rho_{k})\cdot y_{k}+c_{k}\geq0,\label{eq:BITOCPM_b}\\
 & \left\Vert \left[\begin{array}{c}
\textrm{vec}(\mathbf{A}_{k})\\
\sqrt{2}\mathbf{u}_{ce,k}
\end{array}\right]\right\Vert _{2}\leq x_{k},\label{eq:BITOCPM_c}\\
 & y_{k}\mathbf{I}_{MN_{t}}+\mathbf{A}_{k}\succeq\mathbf{0},y_{k}\geq0,\label{eq:BITOCPM_d}\\
 & \textrm{Tr}(\mathbf{G}_{cb}[\mathbf{W}-(\gamma-1)\mathbf{Z}]\mathbf{G}_{cb}^{H}\mathbf{E})\geq(\gamma-1)\sigma_{b}^{2},\label{eq:BITOCPM_bobrate}\\
 & \eqref{eq:OS_powermin_Z}, \eqref{eq:OS_powermin_W}, \eqref{eq:OS_powermin_Wrank1}, \eqref{eq:OS_powermin_unimodulus},
\end{alignat}
\end{subequations}
where $\mathbf{A}=[\mathbf{A}_{1},\mathbf{A}_{2},\cdots,\mathbf{A}_{K}]$. The constraint \eqref{eq:BITOCPM_bobrate} is obtained by substituting \eqref{eq:bobdatarate_c} into \eqref{eq:bobrate}. The constraints \eqref{eq:BITOCPM_b}, \eqref{eq:BITOCPM_c} and \eqref{eq:BITOCPM_d} can be further simplified by some mathematical transformations as follows.

The $\textrm{Tr}(\mathbf{A}_{k})$ in \eqref{eq:BITOCPM_b} can be rewritten as
\begin{equation}
\textrm{Tr}(\mathbf{A}_{k})=\textrm{Tr}(\mathbf{\Sigma}_{e,k}^{1/2}(\mathbf{\Xi}_{e}^{T}\otimes\mathbf{E})\mathbf{\Sigma}_{e,k}^{1/2})=\textrm{Tr}((\mathbf{\Xi}_{e}^{T}\otimes\mathbf{E})\mathbf{\Sigma}_{e,k}).\label{eq:TrAkforoptZW}
\end{equation}

The $\left\Vert \textrm{vec}(\mathbf{A}_{k})\right\Vert _{2}^{2}$
in \eqref{eq:BITOCPM_c} can be written as
\begin{alignat}{1}
\left\Vert \textrm{vec}(\mathbf{A}_{k})\right\Vert _{2}^{2} & =\left\Vert \mathbf{A}_{k}\right\Vert _{F}^{2}=\textrm{Tr}[\mathbf{A}_{k}\mathbf{A}_{k}^{H}]=\textrm{Tr}[\mathbf{\Sigma}_{e,k}^{1/2}(\mathbf{\Xi}_{e}^{T}\otimes\mathbf{E})\mathbf{\Sigma}_{e,k}^{1/2}\mathbf{\Sigma}_{e,k}^{1/2}(\mathbf{\Xi}_{e}^{*}\otimes\mathbf{E}^{H})\mathbf{\Sigma}_{e,k}^{1/2}]\nonumber \\
 & =\textrm{Tr}[(\mathbf{\Xi}_{e}^{T}\otimes\mathbf{E})^{H}\mathbf{\Sigma}_{e,k}(\mathbf{\Xi}_{e}^{T}\otimes\mathbf{E})\mathbf{\Sigma}_{e,k}]\nonumber \\
 & \overset{(c)}{=}\textrm{vec}^{H}(\mathbf{\Xi}_{e}^{T}\otimes\mathbf{E})(\mathbf{\Sigma}_{e,k}^{T}\otimes\mathbf{\Sigma}_{e,k})\textrm{vec}(\mathbf{\Xi}_{e}^{T}\otimes\mathbf{E})\nonumber \\
 & =\textrm{vec}^{H}(\mathbf{\Xi}_{e}^{T}\otimes\mathbf{E})[(\mathbf{\Sigma}_{e,k}^{1/2T}\mathbf{\Sigma}_{e,k}^{1/2T})\otimes(\mathbf{\Sigma}_{e,k}^{1/2}\mathbf{\Sigma}_{e,k}^{1/2})]\textrm{vec}(\mathbf{\Xi}_{e}^{T}\otimes\mathbf{E})\nonumber \\
 & =\textrm{vec}^{H}(\mathbf{\Xi}_{e}^{T}\otimes\mathbf{E})[(\mathbf{\Sigma}_{e,k}^{1/2T}\otimes\mathbf{\Sigma}_{e,k}^{1/2})^{H}(\mathbf{\Sigma}_{e,k}^{1/2T}\otimes\mathbf{\Sigma}_{e,k}^{1/2})]\textrm{vec}(\mathbf{\Xi}_{e}^{T}\otimes\mathbf{E})\nonumber \\
 & =\left\Vert (\mathbf{\Sigma}_{e,k}^{1/2T}\otimes\mathbf{\Sigma}_{e,k}^{1/2})\textrm{vec}(\mathbf{\Xi}_{e}^{T}\otimes\mathbf{E})\right\Vert _{2}^{2},\label{eq:vecAkforoptZW}
\end{alignat}
where the $(c)$ is obtained by invoking the identity $\textrm{Tr}(\mathbf{A}^{H}\mathbf{BCD})=\textrm{vec}^{H}(\mathbf{A})(\mathbf{D}^{T}\otimes\mathbf{B})\textrm{vec}(\mathbf{C})$.

The $\left\Vert \mathbf{u}_{ce,k}\right\Vert _{2}^{2}$ in \eqref{eq:BITOCPM_c}
can be written as
\begin{alignat}{1}
\left\Vert \mathbf{u}_{ce,k}\right\Vert_{2}^{2} & =\mathbf{u}_{ce,k}^{H}\mathbf{u}_{ce,k}\nonumber \\
 & =\textrm{vec}^{H}(\mathbf{E}\mathbf{\bar{\mathbf{G}}}_{ce,k})(\mathbf{\Xi}_{e}^{T}\otimes\mathbf{I}_{M})\mathbf{\Sigma}_{e,k}^{1/2}\mathbf{\Sigma}_{e,k}^{1/2}(\mathbf{\Xi}_{e}^{*}\otimes\mathbf{I}_{M})\textrm{vec}(\mathbf{E}\mathbf{\bar{\mathbf{G}}}_{ce,k})\nonumber \\
 & =\left\Vert \mathbf{\Sigma}_{e,k}^{1/2}(\mathbf{\Xi}_{e}^{T}\otimes\mathbf{I}_{M})\textrm{vec}(\mathbf{E}\mathbf{\bar{\mathbf{G}}}_{ce,k})\right\Vert_{2}^{2}. \label{eq:ucekforoptZW}
\end{alignat}

The constraint $y_{k}\mathbf{I}_{n}+\mathbf{A}_{k}\succeq\mathbf{0},y_{k}\geq0$
in \eqref{eq:BITOCPM_d} can be written as
\begin{equation}
y_{k}\mathbf{I}_{N_{t}M}+\mathbf{\Sigma}_{e,k}^{1/2}(\mathbf{\Xi}_{e}^{T}\otimes\mathbf{E})\mathbf{\Sigma}_{e,k}^{1/2}\succeq\mathbf{0},y_{k}\geq0.\label{eq:eigenforoptZW}
\end{equation}

Finally, by substituting \eqref{eq:TrAkforoptZW},
\eqref{eq:vecAkforoptZW}, \eqref{eq:ucekforoptZW} and \eqref{eq:eigenforoptZW}
into \eqref{BITOCPM}, the OC-PM Problem can be recast as

\begin{subequations}\label{BITOCPMsimplfy}

\begin{alignat}{1}
\underset{\mathbf{W},\mathbf{Z},{\bf \bm{\phi}},\mathbf{x},\mathbf{y}}{\min} & {\rm Tr(}{\bf W}{\rm +}{\bf Z}{\rm )}\label{eq:BITOCPMsimplfypow}\\
\textrm{s.t.}\quad & \textrm{Tr}((\mathbf{\Xi}_{e}^{T}\otimes\mathbf{E})\mathbf{\Sigma}_{e,k})-\sqrt{-2\ln(\rho_{k})}\cdot x_{k}+\ln(\rho_{k})\cdot y_{k}\nonumber \\
 & \;+\textrm{Tr}[\mathbf{\bar{\mathbf{G}}}_{ce,k}\mathbf{\Xi}_{e}\mathbf{\bar{\mathbf{G}}}_{ce,k}^{H}\mathbf{E}]+(\beta-1)\sigma_{e,k}^{2}\geq0,\label{eq:BITOCPMsimplfyLMI}\\
 & \left\Vert \left[\begin{array}{c}
(\mathbf{\Sigma}_{e,k}^{1/2T}\otimes\mathbf{\Sigma}_{e,k}^{1/2})\textrm{vec}(\mathbf{\Xi}_{e}^{T}\otimes\mathbf{E})\\
\sqrt{2}\mathbf{\Sigma}_{e,k}^{1/2}(\mathbf{\Xi}_{e}^{T}\otimes\mathbf{I}_{M})\textrm{vec}(\mathbf{E}\mathbf{\bar{\mathbf{G}}}_{ce,k})
\end{array}\right]\right\Vert _{2}\leq x_{k},\label{eq:BITOCPMsimplfySOC}\\
 & y_{k}\mathbf{I}_{N_{t}M}+\mathbf{\Sigma}_{e,k}^{1/2}(\mathbf{\Xi}_{e}^{T}\otimes\mathbf{E})\mathbf{\Sigma}_{e,k}^{1/2}\succeq\mathbf{0},y_{k}\geq0,\label{eq:BITOCPMsimplfyeig}\\
 & \eqref{eq:BITOCPM_bobrate}, \eqref{eq:OS_powermin_Z}, \eqref{eq:OS_powermin_W}, \eqref{eq:OS_powermin_Wrank1}, \eqref{eq:OS_powermin_unimodulus}.
\end{alignat}
\end{subequations}
\subsection{Solving the Beamforming Matrix and AN Covariance Matrix}
Problem \eqref{BITOCPMsimplfy} is not convex due to the
coupled $\mathbf{\Xi}_{e}$ and $\bm{\phi}$. To solve it, we use
the AO method to update $\left\{ \mathbf{W},\mathbf{Z},\mathbf{x},\mathbf{y}\right\} $
and $\mathbf{\bm{\phi}}$ iteratively. Specifically, when $\mathbf{\bm{\phi}}$
is fixed, Problem \eqref{BITOCPMsimplfy} becomes convex for $\left\{ \mathbf{W},\mathbf{Z},\mathbf{x},\mathbf{y}\right\} $
if the rank one constraint in \eqref{eq:OS_powermin_Wrank1} is relaxed.
By the SDR technique, the problem of optimizing $\left\{ \mathbf{W},\mathbf{Z},\mathbf{x},\mathbf{y}\right\} $
becomes
\begin{subequations}\label{BITOCPMsimplfyWZxy}
\begin{alignat}{1}
\underset{\mathbf{W},\mathbf{Z},\mathbf{x},\mathbf{y}}{\min} & {\rm Tr(}{\bf W}{\rm +}{\bf Z}{\rm )}\\
\textrm{s.t.}\quad & \textrm{Tr}((\mathbf{\Xi}_{e}^{T}\otimes\mathbf{E})\mathbf{\Sigma}_{e,k})-\sqrt{-2\ln(\rho_{k})}\cdot x_{k}+\ln(\rho_{k})\cdot y_{k}\nonumber \\
 & \;+\textrm{Tr}[\mathbf{\bar{\mathbf{G}}}_{ce,k}\mathbf{\Xi}_{e}\mathbf{\bar{\mathbf{G}}}_{ce,k}^{H}\mathbf{E}]+(\beta-1)\sigma_{e,k}^{2}\geq0,\\
 & \left\Vert \left[\begin{array}{c}
(\mathbf{\Sigma}_{e,k}^{1/2T}\otimes\mathbf{\Sigma}_{e,k}^{1/2})\textrm{vec}(\mathbf{\Xi}_{e}^{T}\otimes\mathbf{E})\\
\sqrt{2}\mathbf{\Sigma}_{e,k}^{1/2}(\mathbf{\Xi}_{e}^{T}\otimes\mathbf{I}_{M})\textrm{vec}(\mathbf{E}\mathbf{\bar{\mathbf{G}}}_{ce,k})
\end{array}\right]\right\Vert _{2}\leq x_{k},\\
 & y_{k}\mathbf{I}_{N_{t}M}+\mathbf{\Sigma}_{e,k}^{1/2}(\mathbf{\Xi}_{e}^{T}\otimes\mathbf{E})\mathbf{\Sigma}_{e,k}^{1/2}\succeq\mathbf{0},y_{k}\geq0,\\
 & \eqref{eq:BITOCPM_bobrate}, \eqref{eq:OS_powermin_Z}, \eqref{eq:OS_powermin_W}.
\end{alignat}
\end{subequations}

Problem \eqref{BITOCPMsimplfyWZxy} is convex, and can be solved
by the CVX toolbox. Due to the SDR, the obtained $\mathbf{W}$ of Problem
\eqref{BITOCPMsimplfyWZxy} may not be a rank-one solution. If not,
the suboptimal beamforming vector can be obtained by using the Gaussian
randomization method. Numerical simulations show that the obtained $\mathbf{W}$
always satisfies the rank-one constraint. Thus, the beamforming vector $\mathbf{w}$ can be obtained from the eigen-decomposition of $\mathbf{W}$.

\subsection{Solving the Phase Shifts of IRS}
When $\left\{ \mathbf{W},\mathbf{Z},\mathbf{x},\mathbf{y}\right\} $
are fixed, Problem \eqref{BITOCPMsimplfy} becomes a feasibility
check problem. In order to improve the converged solution in the optimization process,
the data rate inequalities for Bob in \eqref{eq:bobdatarate_c} and
Eve in \eqref{eq:evedatarate_a} are rewritten by introducing slack
variables, and recast respectively as
\begin{subequations}
\begin{align}
 &\quad \quad \quad \quad \textrm{Tr}(\mathbf{G}_{cb}[\mathbf{W}-(\gamma-1)\mathbf{Z}]\mathbf{G}_{cb}^{H}\mathbf{E})\geq(\gamma-1)\sigma_{b}^{2}+\delta_{0},\delta_{0}\geq0, \label{eq:bobraterelaxed}\\
&[\bm{\phi}^{T}(\mathbf{\bar{\mathbf{G}}}_{ce,k}+\triangle\mathbf{G}_{ce,k})]\mathbf{\Xi}_{e}[\bm{\phi}^{T}(\mathbf{\bar{\mathbf{G}}}_{ce,k}+\triangle\mathbf{G}_{ce,k})]^{H}+(\beta-1)\sigma_{e,k}^{2}+\delta_{k}\geq0,\delta_{k}\geq0.\label{eq:everaterelaxed}
\end{align}
\end{subequations}

Then by fixing $\mathbf{W}$ and $\mathbf{Z}$ obtained in the previous
iteration, and using the BTI for the outage of leaked data rate again,
the optimization problem for $\bm{\phi}$ becomes
\begin{subequations}\label{Findfaiform2}
\begin{alignat}{1}
\underset{\bm{\phi},\bm{\delta},\mathbf{x},\mathbf{y}}{\textrm{max}} & \ \ \left\Vert \mathbf{\bm{\delta}}\right\Vert _{1}\\
\textrm{s.t.}\quad & \textrm{Tr}((\mathbf{\Xi}_{e}^{T}\otimes\mathbf{E})\mathbf{\Sigma}_{e,k})-\sqrt{-2\ln(\rho_{k})}\cdot x_{k}+\ln(\rho_{k})\cdot y_{k}\nonumber \\
 & \;+\textrm{Tr}[\mathbf{\bar{\mathbf{G}}}_{ce,k}\mathbf{\Xi}_{e}\mathbf{\bar{\mathbf{G}}}_{ce,k}^{H}\mathbf{E}]+(\beta-1)\sigma_{e,k}^{2}+\delta_{k}\geq0,\label{eq:findfaiform2frst}\\
 & \left\Vert \left[\begin{array}{c}
(\mathbf{\Sigma}_{e,k}^{1/2T}\otimes\mathbf{\Sigma}_{e,k}^{1/2})\textrm{vec}(\mathbf{\Xi}_{e}^{T}\otimes\mathbf{E})\\
\sqrt{2}\mathbf{\Sigma}_{e,k}^{1/2}(\mathbf{\Xi}_{e}^{T}\otimes\mathbf{I}_{M})\textrm{vec}(\mathbf{E}\mathbf{\bar{\mathbf{G}}}_{ce,k})
\end{array}\right]\right\Vert _{2}\leq x_{k},\label{eq:eq:findfaiform2SOC}\\
 & y_{k}\mathbf{I}_{N_{t}M}+\mathbf{\Sigma}_{e,k}^{1/2}(\mathbf{\Xi}_{e}^{T}\otimes\mathbf{E})\mathbf{\Sigma}_{e,k}^{1/2}\succeq\mathbf{0},y_{k}\geq0,\label{eq:findfaiform2eiginequality}\\
 & \textrm{Tr}(\mathbf{G}_{cb}[\mathbf{W}-(\gamma-1)\mathbf{Z}]\mathbf{G}_{cb}^{H}\mathbf{E})\geq(\gamma-1)\sigma_{b}^{2}+\delta_{0},\label{eq:findfaiform2equality}\\
 & \mathbf{\bm{\delta}}\succeq\mathbf{0},\label{eq:deltapostive}\\
 &\eqref{eq:OS_powermin_unimodulus},
\end{alignat}
\end{subequations}
where $\mathbf{\bm{\delta}}=[\delta_{0},\delta_{1},\delta_{2},\cdots,\delta_{K}]^{T}$. Then, the only nonconvex constraint in Problem \eqref{Findfaiform2} is
the unit-modulus constraint \eqref{eq:OS_powermin_unimodulus}. There is no general approach to solve
unit modulus constrained non-convex optimization problems optimally.
To deal with it, the semidefinite programming (SDP) method is utilized. To facilitate SDP, Problem \eqref{Findfaiform2} is recast as
\begin{subequations}\label{Findfaiform2SDP}
\begin{alignat}{1}
\underset{\mathbf{E},\bm{\delta},\mathbf{x},\mathbf{y}}{\textrm{max}} & \ \ \left\Vert \mathbf{\bm{\delta}}\right\Vert _{1}\\
\textrm{s.t.}\quad & \eqref{eq:findfaiform2frst}, \eqref{eq:eq:findfaiform2SOC}, \eqref{eq:findfaiform2eiginequality}, \eqref{eq:findfaiform2equality}, \eqref{eq:deltapostive}\\
 & \textrm{Diag(\ensuremath{\mathbf{E}})}=\mathbf{1}_{M}, \label{eq:findfaiform2SDPdiag}\\
 &\mathbf{E}\succeq\mathbf{0}, \label{eq:findfaiform2SDPsemi}\\
 &  \textrm{rank}(\mathbf{E})=1,\label{eq:findfaiform2SDPrank1}
\end{alignat}
\end{subequations}
where the constraints \eqref{eq:findfaiform2SDPdiag}, \eqref{eq:findfaiform2SDPsemi} and \eqref{eq:findfaiform2SDPrank1} are equivalent to the constraint \eqref{eq:OS_powermin_unimodulus}, and they are imposed to ensure that the
phase shifts vector $\bm{\phi}$ with unit modulus can be recovered
from $\mathbf{E}$. The Problem \eqref{Findfaiform2SDP} is convex except the nonconvex constraint \eqref{eq:findfaiform2SDPrank1}. Although the SDR method can be used to solve Problem \eqref{Findfaiform2SDP} by removing constraint \eqref{eq:findfaiform2SDPrank1} and solving the resulted SDP problem, the rank of solved $\mathbf{E}$ is generally larger than one. To handle this problem,
we construct a convex constraint equivalent to the rank one constraint \eqref{eq:findfaiform2SDPrank1}, and the following Lemma is utilized.
\newtheorem{lemma2}{Lemma}
\begin{lemma}\label{Lemma_rankone}
For any positive semidefinite matrix $\mathbf{A}$, the following
inequality holds
\begin{alignat}{1}
\left|\mathbf{I}+\mathbf{A}\right| & \geq1+\text{{Tr}}(\mathbf{A}),\label{eq:rankoneinequality}
\end{alignat}
and the equality in \eqref{eq:rankoneinequality} holds if and only
if $\textrm{rank}(\mathbf{A})\leq1$.
\emph{Proof}: \upshape
Let $r_{A}=\textrm{rank}(\mathbf{A})$. Since $r_{A}=0$ is
trivial, the nonzero eigenvalues of $\mathbf{A}$ are denoted by $\lambda_{1}\geq\lambda_{2}\geq\cdots\geq\lambda_{r_{A}}>0$.
Then we have
\begin{alignat}{1}
\left|\mathbf{I}+\mathbf{A}\right| & =\prod_{i=1}^{r_{A}}(1+\lambda_{i})=1+\sum\limits _{i=1}^{r_{A}}\lambda_{i}+\sum\limits _{i\neq k}\lambda_{i}\lambda_{k}+\cdots\nonumber \\
 & \geq1+\sum\limits _{i=1}^{r_{A}}\lambda_{i}=1+\text{{Tr}}(\mathbf{A}).\label{eq:rankoneinequalityexpand}
\end{alignat}
It can be seen from \eqref{eq:rankoneinequalityexpand} that the equality
holds if and only if $\textrm{rank}(\mathbf{A})=1$.
\hfill\rule{2.7mm}{2.7mm}

\end{lemma}

By invoking Lemma \ref{Lemma_rankone}, we can build an equivalent
constraint as
\begin{subequations}
\begin{alignat}{1}
\eqref{eq:findfaiform2SDPrank1}\Leftrightarrow & \left|\mathbf{I}+\mathbf{E}\right|\leq1+\textrm{Tr}(\mathbf{E}), \\
\Leftrightarrow & \log\det(\mathbf{I}+\mathbf{E})\leq\log(1+M), \label{eq:rankoneequivltconstr}
\end{alignat}
\end{subequations}
where $\textrm{Tr}(\mathbf{E})=M$. The constraint \eqref{eq:rankoneequivltconstr} ensures the rank one
equality, thus is equivalent to constraint \eqref{eq:findfaiform2SDPrank1}. By utilizing the penalty-based method and putting the constraint \eqref{eq:rankoneequivltconstr} into the objective function (OF) of Problem
\eqref{Findfaiform2SDP}, Problem
\eqref{Findfaiform2SDP} can be cast as
\begin{subequations}\label{Findfaiform2SDPpenalty}
\begin{alignat}{1}
\underset{\mathbf{E},\bm{\delta},\mathbf{x},\mathbf{y}}{\textrm{max}} & \ \ \left\Vert \mathbf{\bm{\delta}}\right\Vert _{1}+\kappa[\log(1+M)-\log\det(\mathbf{I}+\mathbf{E})]\\
\textrm{s.t.}\quad & \eqref{eq:findfaiform2frst}, \eqref{eq:eq:findfaiform2SOC}, \eqref{eq:findfaiform2eiginequality}, \eqref{eq:findfaiform2equality}, \eqref{eq:deltapostive}, \eqref{eq:findfaiform2SDPdiag}, \eqref{eq:findfaiform2SDPsemi},
\end{alignat}
\end{subequations}
where $\kappa$ is a penalty factor penalizing the violation of constraint
$\textrm{rank}(\mathbf{E})=1$. Since the $\log\det(\mathbf{I}+\mathbf{E})$
is a concave function with respect to $\mathbf{E}$, the upper bound of it can
be obtained by using the first-order Taylor approximation as
\begin{alignat}{1}
\log\det(\mathbf{I}+\mathbf{E})\leq & (\log e)\textrm{Tr}\{((\mathbf{I}+\mathbf{E}^{(n)})^{-1})^{*}(\mathbf{E}-\mathbf{E}^{(n)})\}+(\log e)\log\det(\mathbf{I}+\mathbf{E}^{(n)}),\label{eq:rankoneequivltconstrtaylor}
\end{alignat}
where $e$ denotes natural logarithm, and the $\frac{\partial\ln(\left|\textrm{det}(\mathbf{X})\right|)}{\partial\mathbf{X}}=(\mathbf{X}^{-1})^{T}$ is
utilized. By substituting \eqref{eq:rankoneequivltconstrtaylor} into the OF of Problem
\eqref{Findfaiform2SDPpenalty} and removing the constant terms, Problem
\eqref{Findfaiform2SDPpenalty} can be written as
\begin{subequations}\label{Findfaiform2SDPpenaltycvx}
\begin{alignat}{1}
\underset{\mathbf{E},\bm{\delta},\mathbf{x},\mathbf{y}}{\textrm{max}} & \ \ \left\Vert \mathbf{\bm{\delta}}\right\Vert _{1}+\kappa[-(\log e)\textrm{Tr}\{((\mathbf{I}+\mathbf{E}^{(n)})^{-1})^{*}(\mathbf{E}-\mathbf{E}^{(n)})\}]\\
\textrm{s.t.} \quad & \eqref{eq:findfaiform2frst}, \eqref{eq:eq:findfaiform2SOC}, \eqref{eq:findfaiform2eiginequality}, \eqref{eq:findfaiform2equality}, \eqref{eq:deltapostive}, \eqref{eq:findfaiform2SDPdiag}, \eqref{eq:findfaiform2SDPsemi}.
\end{alignat}
\end{subequations}

Problem \eqref{Findfaiform2SDPpenaltycvx} is jointly convex
with respect to $\{\mathbf{E},\bm{\delta},\mathbf{x},\mathbf{y}\}$,
hence it can be efficiently solved by standard convex program solvers
such as CVX. A rank-one solution $\mathbf{E}^{\star}$
can be obtained by solving Problem \eqref{Findfaiform2SDPpenaltycvx} for a sufficiently
small value of penalty factor $\kappa$. The maximum value of Problem \eqref{Findfaiform2SDPpenaltycvx}
serves as the lower bound for the optimal value of Problem \eqref{Findfaiform2SDPpenalty}.
The overall AO algorithm proposed is summarized in Algorithm \ref{AO}. By iteratively solving Problem \eqref{BITOCPMsimplfyWZxy}
and Problem \eqref{Findfaiform2SDPpenaltycvx} optimally in
Step 3 and Step 4 in Algorithm \ref{AO}, the transmit power can be monotonically
reduced with guaranteed convergence.
\begin{algorithm}
\caption{Alternating Optimization Algorithm}
\label{AO} \begin{algorithmic}[1] \STATE Parameter Setting:
Set the maximum number of iterations $n_{{\rm {max}}}$ and the first
iterative number $n=1$; Give the penalty factor $\kappa$ and error
tolerance $\varepsilon$;

\STATE Initialize the variables ${\bf {w}}^{(1)}$, ${\bf {Z}}^{(1)}$
and ${\bm{\phi}}^{(1)}$ in the feasible region; Compute the OF value
of Problem \eqref{BITOCPMsimplfy} as ${\rm {OF(}}{{\bf {w}}^{(1)}},{{\bf {Z}}^{(1)}}{\rm {)}}$;

\STATE Solve Problem \eqref{BITOCPMsimplfyWZxy} to obtain the ${\bf {w}}^{(n)},{{\bf {Z}}^{(n)}}$
by fixing ${\bm{\phi}}^{(n-1)}$; Calculate the OF value of Problem
\eqref{BITOCPMsimplfyWZxy} as ${\rm {OF(}}{{\bf {w}}^{(n)}},{{\bf {Z}}^{(n)}}{\rm {)}}$;

\STATE Solve Problem \eqref{Findfaiform2SDPpenaltycvx} to obtain
the ${\bm{\phi}}^{(n)}$ by fixing ${\bf {w}}^{(n)},{{\bf {Z}}^{(n)}}$;

\STATE If ${{\left|\!{{\rm {OF}}\!(\!{\bf {w}}^{(n)}\!,\!{{\bf {Z}}^{(n)}}\!)\!\!-\!\!{\rm {OF}}\!(\!{\bf {w}}^{(n-1)}\!,\!{{\bf {Z}}^{(n-1)}}\!\!)}\!\right|}/\!{{\rm {OF}}\!(\!{\bf {w}}^{(n-1)}\!,\!{{\bf {Z}}^{(n-1)}}\!\!)}}\!<\varepsilon$
or $n\geq n_{{\rm {max}}}$, terminate. Otherwise, update $n\leftarrow n+1$
and jump to Step 3. \end{algorithmic}
\end{algorithm}

\section{Simulation Results}

\subsection{Simulation Setup}

\begin{figure}
\centering\includegraphics{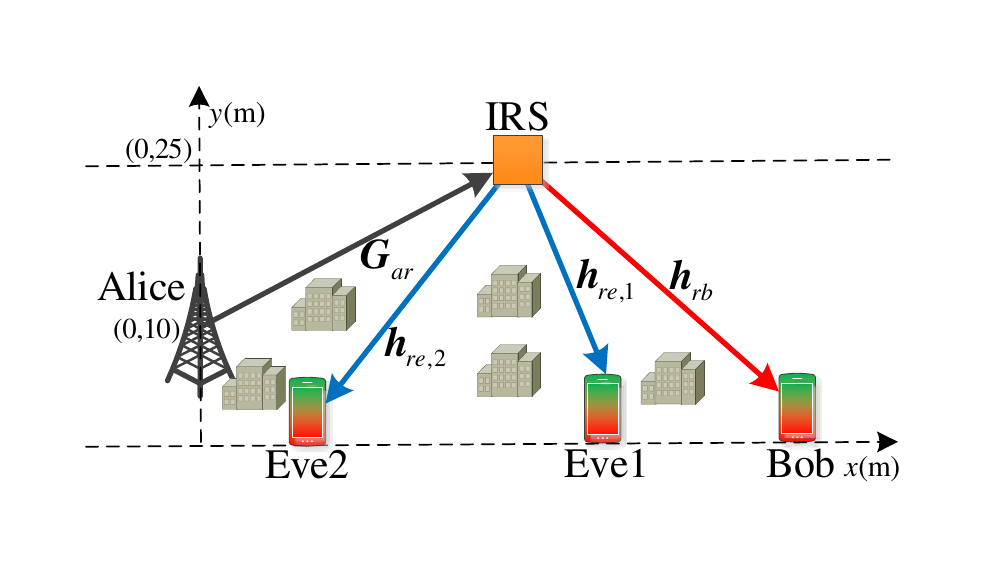}

\caption{The simulated secure communication scenario}

\label{figsimulscenario}
\end{figure}

One appealing benefit of deploying an IRS in secure wireless systems is
to establish favorable communication links for legitimate user that
is blocked by obstacles. We consider a scenario where the direct links
from the AP to the legitimate user and evesdroppers are blocked. The
simulated system model is shown in Fig. \ref{figsimulscenario}.
The Alice is located at (0,10) m. The IRS is installed at a height of 25 m. The Bob and
Eves can be randomly distributed on the x axis.

The channel matrix
between the Alice and the IRS is modeled as $\mathbf{\mathbf{G}}_{ar}\in\mathbb{\mathbb{C}}^{M\times N_{t}}$, which is
\begin{alignat}{1}
\mathbf{\mathbf{G}}_{ar}^{T} & =\sqrt{L_{0}d_{ar}^{-\alpha_{ar}}}\left(\sqrt{\frac{\beta_{ar}}{1+\beta_{ar}}}\mathbf{G}_{ar}^{LOS}+\sqrt{\frac{1}{1+\beta_{ar}}}\mathbf{G}_{ar}^{NLOS}\right),
\end{alignat}
where $L_{0}=\left(\frac{\lambda_{c}}{4\pi}\right)^{2}$ is a constant
with $\lambda_{c}$ being the wavelength of the center frequency of
the information carrier. The distance between the AP and the IRS is denoted
by $d_{ar}$, while $\alpha_{ar}$ denotes the corresponding path loss
exponent. The small-scale fading is assumed to be Ricean fading with
Ricean factor $\beta_{ar}$. The $\mathbf{G}_{ar}^{LOS}$ denotes the light of sight (LoS) component of the Alice-IRS channel. By assuming that the antennas at
the Alice and the passive reflecting elements at the IRS are both arranged
in a uniform linear array (ULA), the $\mathbf{G}_{ar}^{LOS}$ can
be modeled as $\mathbf{G}_{ar}^{LOS}=\mathbf{c}\mathbf{d}^{H}$, where
$\mathbf{c}=[c_{1},c_{2},\cdots,c_{N_{t}}]^{T}$ denotes the transmit steering
vector of the Alice, and $\mathbf{d}=[d_{1},d_{2},\cdots,d_{M}]^{T}$
denotes the receive steering vector of the IRS. The $m$th elements
of $\mathbf{c}$ and $\mathbf{d}$ are written as
\begin{subequations}
\begin{alignat}{1}
c_{m} & =\exp(j2\pi\frac{d_{a}}{\lambda_{c}}(m-1)\sin\theta_{a}),\\
d_{m} & =\exp(j2\pi\frac{d_{r}}{\lambda_{c}}(m-1)\sin\theta_{r}),
\end{alignat}
\end{subequations}
where $d_{a}$ and $d_{r}$ are respectively the element intervals
at Alice and the IRS; $\theta_{a}$ represents the elevation angle of
the LoS link from Alice; $\theta_{r}$ represents the elevation
angle of the LoS link to the IRS. We set $\frac{d_{a}}{\lambda_{c}}=\frac{d_{r}}{\lambda_{c}}=0.5$,
$\theta_{a}=\tan^{-1}(\frac{y_{r}-y_{a}}{x_{r}-x_{a}})$, and $\theta_{r}=\pi-\theta_{a}$.
$\mathbf{G}_{ar}^{NLOS}$ denotes the non-LoS (NLoS) component of the Alice-IRS
channel, which is modeled as Rayleigh fading. The channels of the
IRS-Bob link and IRS-Eve link are modeled similarly, where both the
LoS component and NLoS component exist simultaneously.

\begin{table} \centering
\caption{Simulation Parameters}

\begin{tabular}{|>{\centering}p{8cm}|c|}
\hline
Carrier center frequency  & 2.4GHz\tabularnewline
\hline
Path loss exponents for Alice-IRS channels $\alpha_{ar}$, IRS-Bob channels $\alpha_{rb}$, IRS-Eve channels $\alpha_{re}$,
  & 2\tabularnewline
\hline
Rician factor for Alice-IRS channel, $\beta_{ar}$  & 5\tabularnewline
\hline
Rician factor for IRS-Bob channel, $\beta_{rb}$  & 5\tabularnewline
\hline
Rician factor for IRS-Eve channel, $\beta_{re}$  & 5\tabularnewline
\hline
Noise power at Bob and Eves, $\sigma_{b}^{2}$, $\sigma_{e,k}^{2}$  & -75dBm\tabularnewline
\hline
Penalty factor, $\kappa$  & $\textrm{5}\times10^{-8}$\tabularnewline
\hline
Convergence tolerance, $\varepsilon$  & $10^{-3}$\tabularnewline
\hline
Outage probability, $\rho_{k}$  & 0.05\tabularnewline
\hline
\end{tabular} \label{tableparamtrs}
\end{table}

For the statistical cascaded CSI error model, the variance matrix of $\mathbf{g}_{ce,k}=\textrm{vec}(\triangle\mathbf{G}_{ce,k})$
is defined as $\mathbf{\Sigma}_{e,k}=\varepsilon_{g,k}^{2}\mathbf{I}$,
where $\varepsilon_{g,k}^{2}=\delta_{g,k}^{2}\left\Vert \textrm{vec}(\mathbf{\bar{\mathbf{G}}}_{ce,k})\right\Vert _{2}^{2}$.
$\delta_{g,k}^{2}\in\left[0,1\right)$ is the normalized CSI error,
which measures the relative amount of CSI uncertainties. The system
parameters used in the following simulations are listed in Table \ref{tableparamtrs}.

\subsection{Benchmark Schemes}

We demonstrate the advantage of the proposed AO algorithm by comparing its performance with the following benchmark schemes:
\begin{itemize}
\item Random-MRT: It performs maximum ratio transmission (MRT) based beamforming
design, i.e., $\mathbf{w}=\sqrt{p_{w}}\frac{\hat{\mathbf{h}}_{b}}{\left\Vert \hat{\mathbf{h}}_{b}\right\Vert }=\sqrt{p_{w}}\frac{\mathbf{G}_{ar}^{H}\mathbf{\Phi}^{H}\mathbf{\mathbf{h}}_{rb}}{\left\Vert \hat{\mathbf{h}}_{b}\right\Vert }$,
where $p_{w}$ is the power allocated to legitimate user Bob. It applies
an isotropic AN\cite{liao2010qos}, i.e., the AN covariance matrix
is chosen as $\mathbf{Z}=p_{z}\mathbf{P}_{\hat{\mathbf{h}}_{b}}^{\perp}$,
where $\mathbf{P}_{\hat{\mathbf{h}}_{b}}^{\perp}=\mathbf{I}_{N_{t}}-\hat{\mathbf{h}}_{b}\hat{\mathbf{h}}_{b}^{H}/\left\Vert \hat{\mathbf{h}}_{b}\right\Vert ^{2}$
is the orthogonal complement projector of $\hat{\mathbf{h}}_{b}$,
and $p_{z}$ is the power invested on AN. Both the beamforming precoder
$\mathbf{w}$ and the AN covariance matrix $\mathbf{Z}$ rely on the
phase shifts $\mathbf{\Phi}$ of IRS. For Random-MRT schemes, we assume
the phase shifts $\mathbf{\Phi}$ of IRS is randomly chosen. The allocated
power $p_{w}$ to beamforming and the allocated power $p_{z}$ to
AN are optimized to satisfy the secrecy constraints in Problem
\eqref{BITOCPMsimplfyWZxy}.
\item Optimized-MRT: Based on the MRT beamforming and isotropic AN, the
phase shifts $\mathbf{\Phi}$ of IRS is obtained by using the proposed method. The allocated power $p_{w}$ to beamforming and
the allocated power $p_{z}$ to AN are optimized to satisfy the secrecy
constraints in Problem \eqref{BITOCPMsimplfyWZxy}.
\item Random-IRS: It does not optimize the phase shifts of IRS, and the
phase shifts of IRS $\mathbf{\Phi}$ is randomly selected. The beamforming
matrix $\mathbf{W}$ and AN covariance matrix $\mathbf{Z}$ are optimized
by solving Problem \eqref{BITOCPMsimplfyWZxy}.
\end{itemize}
The Random-MRT and Optimized-MRT schemes only exploit the CSI of Bob by assuming that the CSI of Eves is unknown, while the Random-IRS scheme and the proposed algorithm exploit the CSI of both Bob and Eves to realize robust transmission design. The Optimized-MRT is compared with the
Random-MRT to demonstrate the importance of the optimization of IRS phase shifts. For Random-IRS scheme, the phase shifts of IRS $\mathbf{\Phi}$
are not optimized, and only the beamforming matrix $\mathbf{W}$ and
AN covariance matrix $\mathbf{Z}$ are optimized robustly, while the proposed algorithm optimizes these variables jointly. By comparing the proposed algorithm with the Random-IRS scheme, the importance of the optimization of the IRS phase shifts in robust transmission design is verified.
%\centering
\begin{figure} \centering
\subfigure[Convergence of the proposed algorithm with different values of $N_t$.] { \label{figPowvsIterfordiffNt}
\includegraphics[width=3.5in]{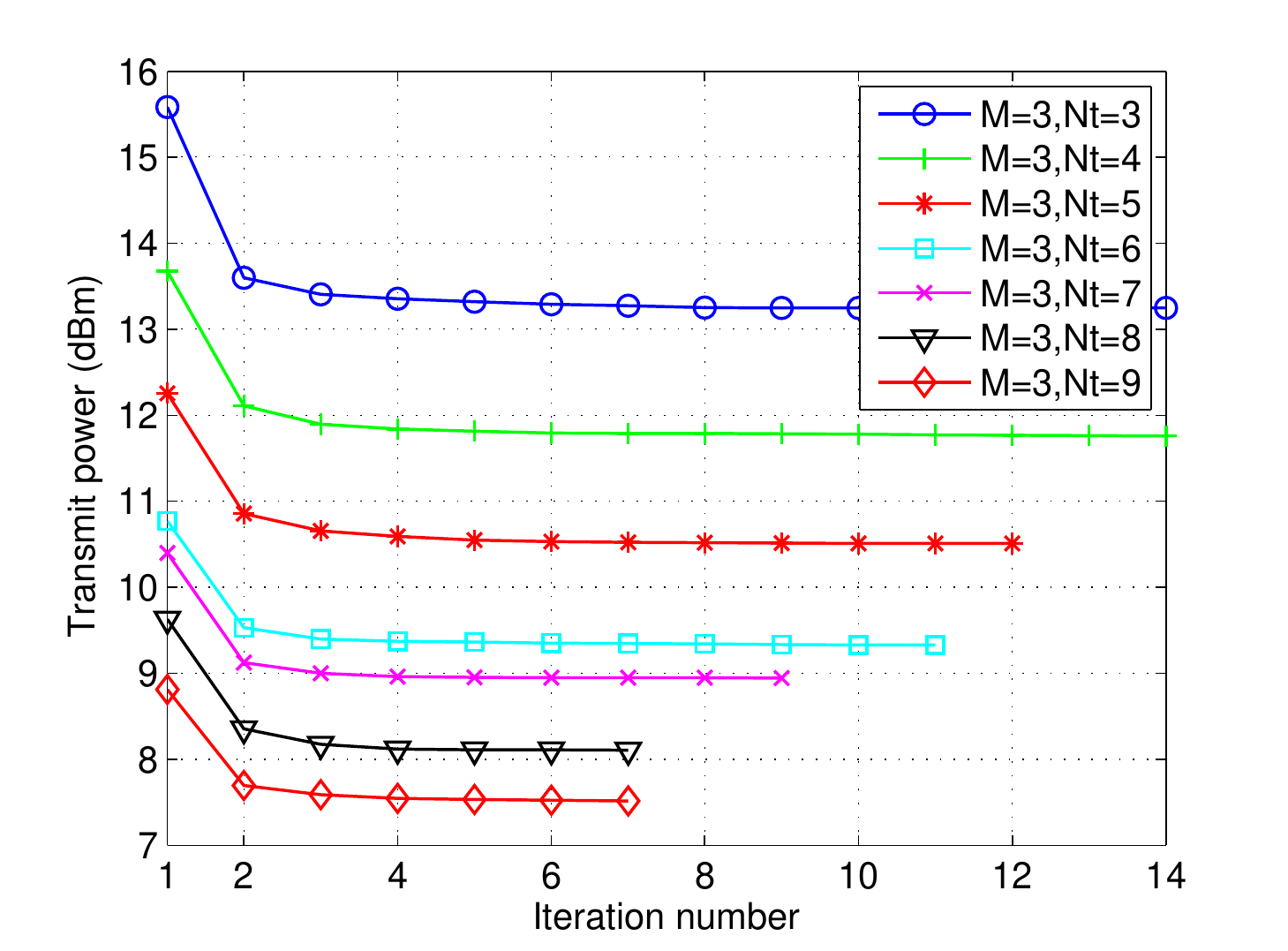}
}
\subfigure[Convergence of the proposed algorithm with different values of $M$.] { \label{figPowvsIterfordiffM}
\includegraphics[width=3.5in]{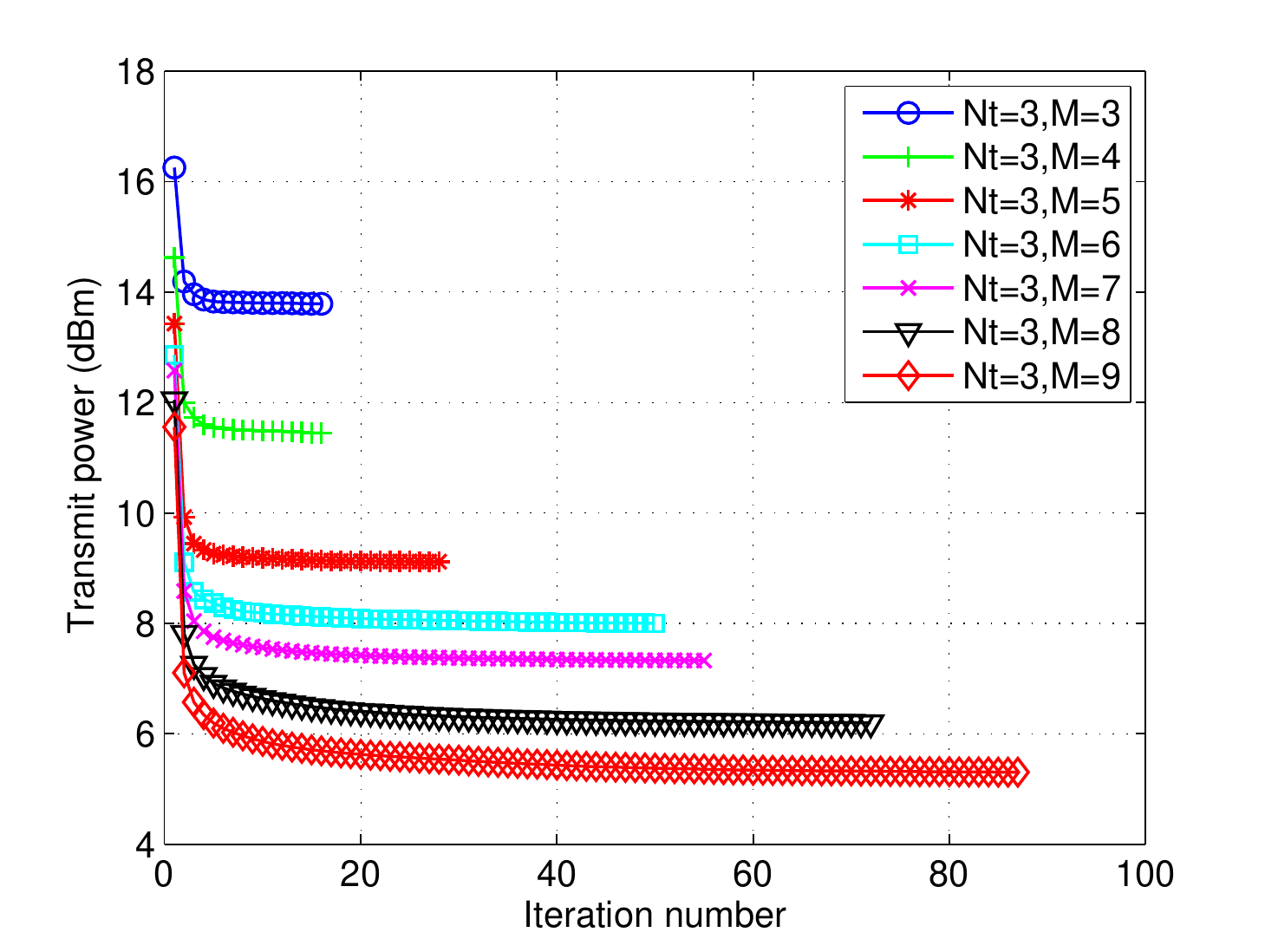}
}
\caption{Convergence of the proposed algorithm with $\log \gamma=3$ $\textrm{bit/s/Hz}$ and $\log\beta=1$ $\textrm{bit/s/Hz}$, $K=2$, $\delta_{g,k}^{2}=0.0001$, $\forall k$. The location of AP is $\textrm{(0, 10) m}$,
the location of IRS is $\textrm{(100, 25) m}$, the location of
Bob is $\textrm{(180, 0) m}$, and the locations of Eves are $\textrm{(160, 0) m}$
and $\textrm{(170, 0) m}$.}
\label{figPowvsIterfordiffMandNt}
\end{figure}

\begin{table}[htbp]
\caption{Time Cost}
\begin{center}
\begin{tabular}{|c|c|c|c|c|c|c|c|}
\hline
\text{Versus}&\multicolumn{7}{|c|}{\text{Average CPU time (secs) in Fig. \ref{figPowvsIterfordiffNt}}} \\
\cline{2-8}
$M$ & $M=3$ & $M=4$ & $M=5$ & $M=6$ & $M=7$ & $M=8$ & $M=9$ \\
\hline
$N_t=3$ & 2.0932& 2.9118& 6.5124& 9.4720& 13.1437& 18.5433& 20.9004  \\
\hline
\text{Versus}&\multicolumn{7}{|c|}{\text{Average CPU time (secs) in Fig. \ref{figPowvsIterfordiffM}}} \\
\cline{2-8}
$N_t$ & $N_t=3$ & $N_t=4$ & $N_t=5$ & $N_t=6$ & $N_t=7$ & $N_t=8$ & $N_t=9$ \\
\hline
$M=3$ & 1.6999& 1.7960& 1.9247& 1.9806& 2.2405& 2.3101& 2.7319  \\
\hline
\text{Versus}&\multicolumn{7}{|c|}{\text{Average CPU time (secs) in Fig. \ref{figPowvsIterfordiffgamma}}} \\
\cline{2-8}
\text{log}$\gamma$ (\text{bit/s/Hz}) & \text{log}$\gamma$=5.5 & \text{log}$\gamma$=5 & \text{log}$\gamma$=4.5 & \text{log}$\gamma$=4 & \text{log}$\gamma$=3.5 & \text{log}$\gamma$=3 & \text{log}$\gamma$=2.5 \\
\hline
\text{log}$\beta$=2 (\text{bit/s/Hz}) & 25.7724& 26.7880& 29.8244& 36.3530& 44.6590& 67.7409& 97.2034  \\
\hline
\text{Versus}&\multicolumn{7}{|c|}{\text{CPU time (secs) in Fig. \ref{figPowvsIterfordiffbeta}}} \\
\cline{2-8}
\text{log}$\beta$ (\text{bit/s/Hz}) & \text{log}$\beta$=1.5 & \text{log}$\beta$=2 & \text{log}$\beta$=2.5 & \text{log}$\beta$=3 & \text{log}$\beta$=3.5 & \text{log}$\beta$=4 & \text{log}$\beta$=4.5 \\
\hline
\text{log}$\gamma$=4.7 (\text{bit/s/Hz}) & 24.4419& 28.2168& 30.1735& 37.1616& 48.4290& 66.7990& 90.5877  \\
\hline
%\multicolumn{4}{l}{$^{\mathrm{a}}$Sample of a Table footnote.}
\end{tabular}
\label{tab1}
\end{center}\label{timecost}
\end{table}
\subsection{Convergence Analysis}
The convergence of the proposed method with different antenna numbers of transmitter and different numbers of IRS phase shifts are investigated in Fig. \ref{figPowvsIterfordiffMandNt}. As observed from Fig. \ref{figPowvsIterfordiffNt}, the proposed algorithm is likely to converge with fewer iterations when the number of transmit antennas increases. With increased $N_t$, the dimension of optimization variables $\mathbf{W}$ and $\mathbf{Z}$ in the CVX problem of each iteration becomes large, which requires more computation time for each iteration. Thus, the computation time consumed by the proposed algorithm with a larger $N_t$ still increases even with fewer iterations. Fig. \ref{figPowvsIterfordiffM} shows the convergence of the proposed algorithm with different values of $M$. It is seen that the proposed algorithm is likely to converge with more iterations when the number of IRS elements increases. Since the quality of the cascaded channel relies on the phase shifts of the IRS, larger $M$ will bring more degrees of freedom to adjust the cascaded channel, thus more iterations are required for the fine adjustment. With increased $M$, the dimension of the optimization variable $\mathbf{\Phi}$ in the CVX problem of each iteration becomes large, thus the computation time of the proposed algorithm increases with $M$. The average central processing unit (CPU) running time of the proposed algorithm versus $N_t$ and $M$  with the parameters in Fig. \ref{figPowvsIterfordiffNt} and Fig. \ref{figPowvsIterfordiffM} is shown in Table \ref{timecost}. The data is obtained by using a computer with a 3.40GHz i7-6700 CPU and 16GB RAM. It is observed from Table \ref{timecost} that the average CPU running time required by our proposed algorithm increases with either $N_t$ or $M$. The average CPU time increases with $N_t$ less slowly than with $M$, this is because the required iterations decrease with $N_t$.
\begin{figure} \centering
\subfigure[Convergence of the proposed algorithm for different values of $\log \gamma$ for $\log\beta=2$ $\textrm{bit/s/Hz}$.] { \label{figPowvsIterfordiffgamma}
\includegraphics[width=3.5in]{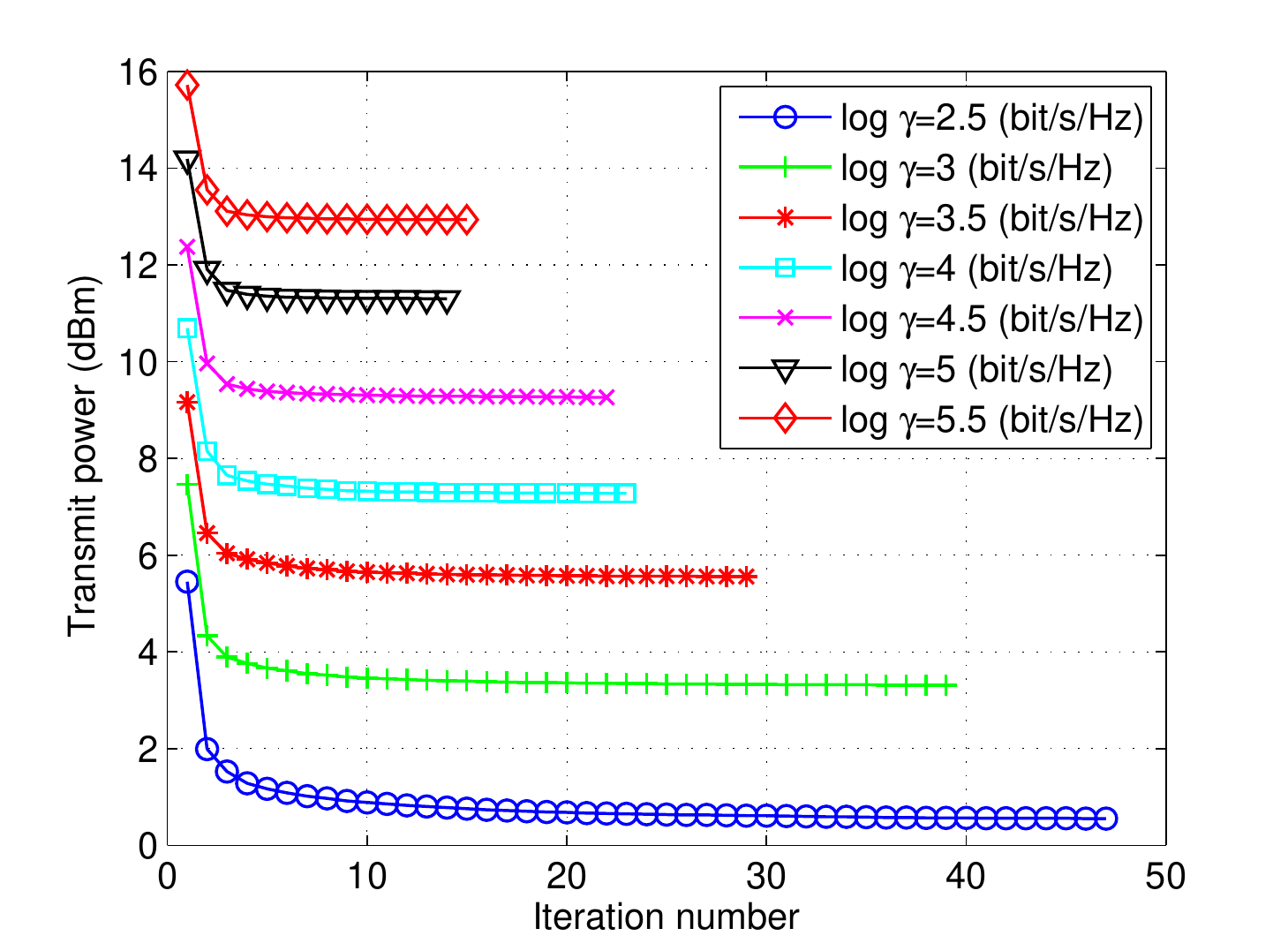}
}
\subfigure[Convergence of the proposed algorithm for different values of $\log \beta$ for $\log\gamma=4.7$ $\textrm{bit/s/Hz}$.] { \label{figPowvsIterfordiffbeta}
\includegraphics[width=3.5in]{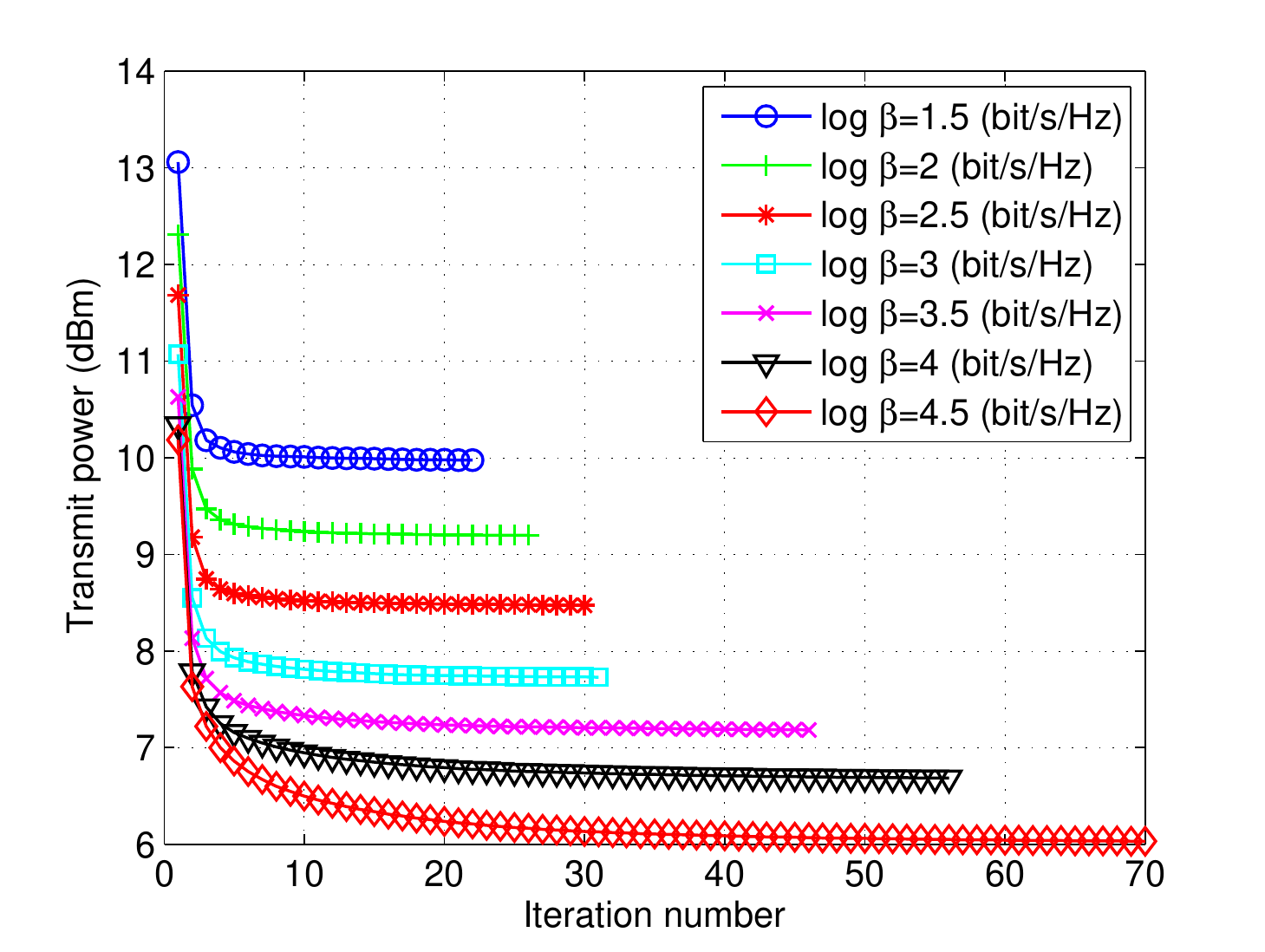}
}
\caption{Convergence of the proposed algorithm with $N_{t}=M=6$, $K=2$, $\delta_{g,k}^{2}=0.0001$, $\forall k$. The location of AP is $\textrm{(0, 10) m}$,
the location of IRS is $\textrm{(50, 25) m}$, the location of
Bob is $\textrm{(180, 0) m}$, and the locations of Eves are $\textrm{(20, 0) m}$
and $\textrm{(10, 0) m}$.}
\label{figPowvsIterfordiffgammaandbeta}
\end{figure}
The convergence of the proposed algorithm for different values of $\log \gamma$ and $\log\beta$ is shown in Fig. \ref{figPowvsIterfordiffgammaandbeta}. It is demonstrated that more iterations are required by decreasing the minimum channel capacity $\log \gamma$ of the legitimate user and increasing the maximum tolerable channel capacity $\log \beta$ of the eavesdroppers. Since the dimension of the CVX problem in each iteration is not changed with different $\log \gamma$ and $\log\beta$, more iterations in Fig. \ref{figPowvsIterfordiffgammaandbeta} signify more computation time. The average CPU running time of the proposed algorithm versus $\log \gamma$ and $\log\beta$ with the parameters in Fig. \ref{figPowvsIterfordiffgamma} and Fig. \ref{figPowvsIterfordiffbeta} is shown in Table \ref{timecost}. It is found in Table \ref{timecost} that more CPU time is required for the small value of secrecy rate.
\subsection{Transmit Power Versus the Minimum Channel Capacity of the Legitimate User}
\begin{figure}[h!]
\centering \includegraphics[width=3.5in]{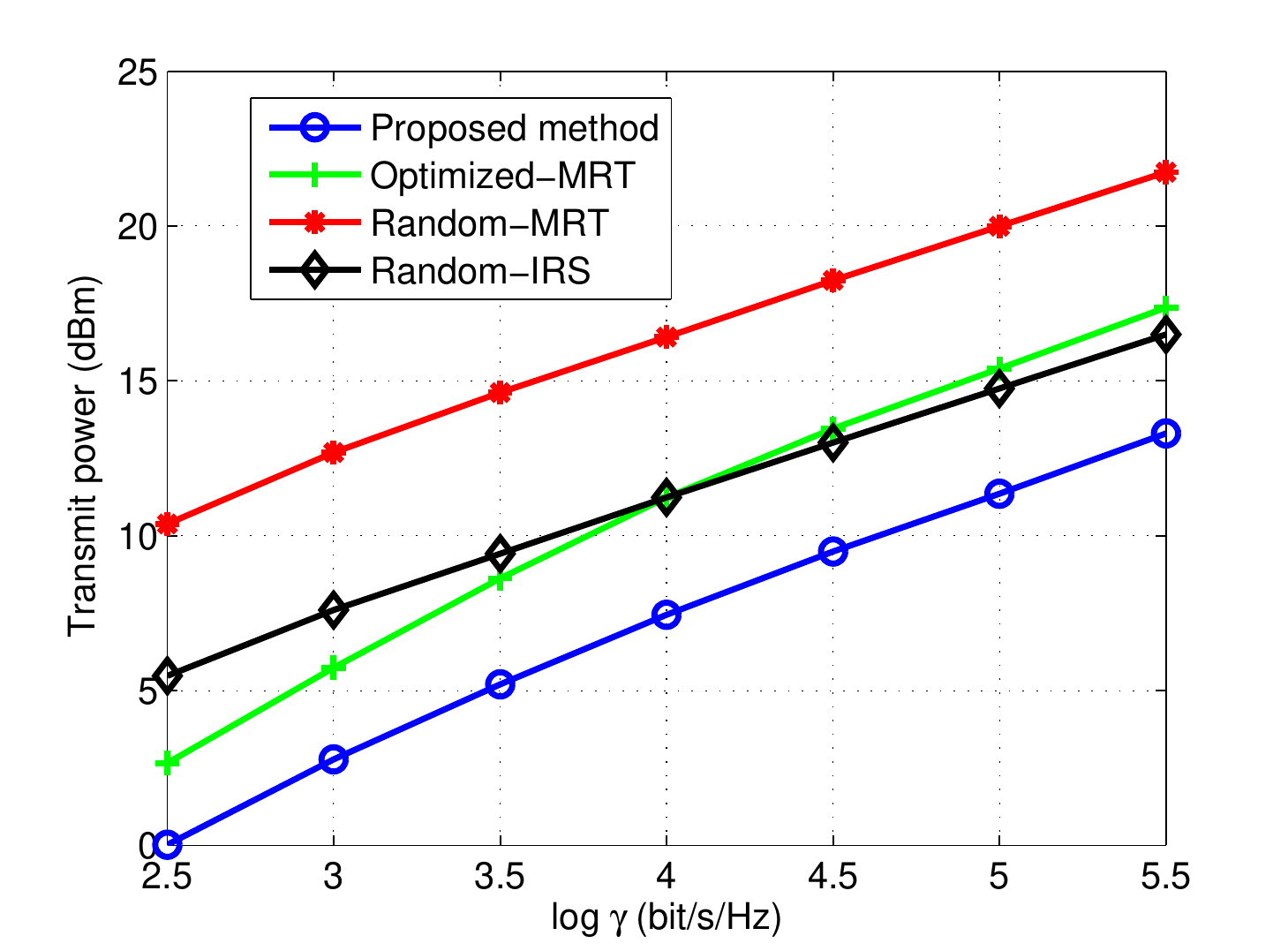}\vspace{-0.2cm}
 \caption{Transmit power versus the minimum channel capacity $\log\gamma$ of
the legitimate user with $N_{t}=M=6$, $K=2$, and $\delta_{g,k}^{2}=0.0001$,
$\forall k$. The location of AP is $\textrm{(0, 10) m}$,
the location of IRS is $\textrm{(50, 25) m}$, the location of
Bob is $\textrm{(180, 0) m}$, and the locations of Eves are $\textrm{(20, 0) m}$
and $\textrm{(10, 0) m}$.}\vspace{-0.5cm}
\label{figPvsBobGamma}
\end{figure}

In Fig. \ref{figPvsBobGamma}, we show the transmit power versus the
minimum channel capacity $\log\gamma$ of the legitimate user, while
limiting the information leakage to the potential eavesdroppers as
$\log\beta=2$ $\textrm{bit/s/Hz}$. From Fig. \ref{figPvsBobGamma},
we find that the transmit power increases monotonically with the minimum
channel capacity of the Bob. This is because more transmit power is
required to satisfy the increased data rate of Bob. By comparing our
proposed method with the other benchmark schemes, it is seen that the power budget of the proposed method
is lower than the other benchmark schemes.
In particular, the MRT schemes only exploit the channel information
of Bob, thus their power consumptions stay at a high level. The optimized
MRT scheme is better than the random MRT scheme, because the IRS-relied
cascaded channel has been greatly improved by optimizing the phase
shifts of IRS. The Random-IRS scheme consumes more power than the
proposed method, which demonstrates the necessity of optimizing the phase shifts of IRS. Fig. \ref{figPvsBobGamma} reveals that the
security communication in LoS blocked environment can be realized
by deploying an IRS, and the phase shifts of IRS should be optimized.

\begin{figure}[h!]
\centering
\includegraphics[width=3.5in]{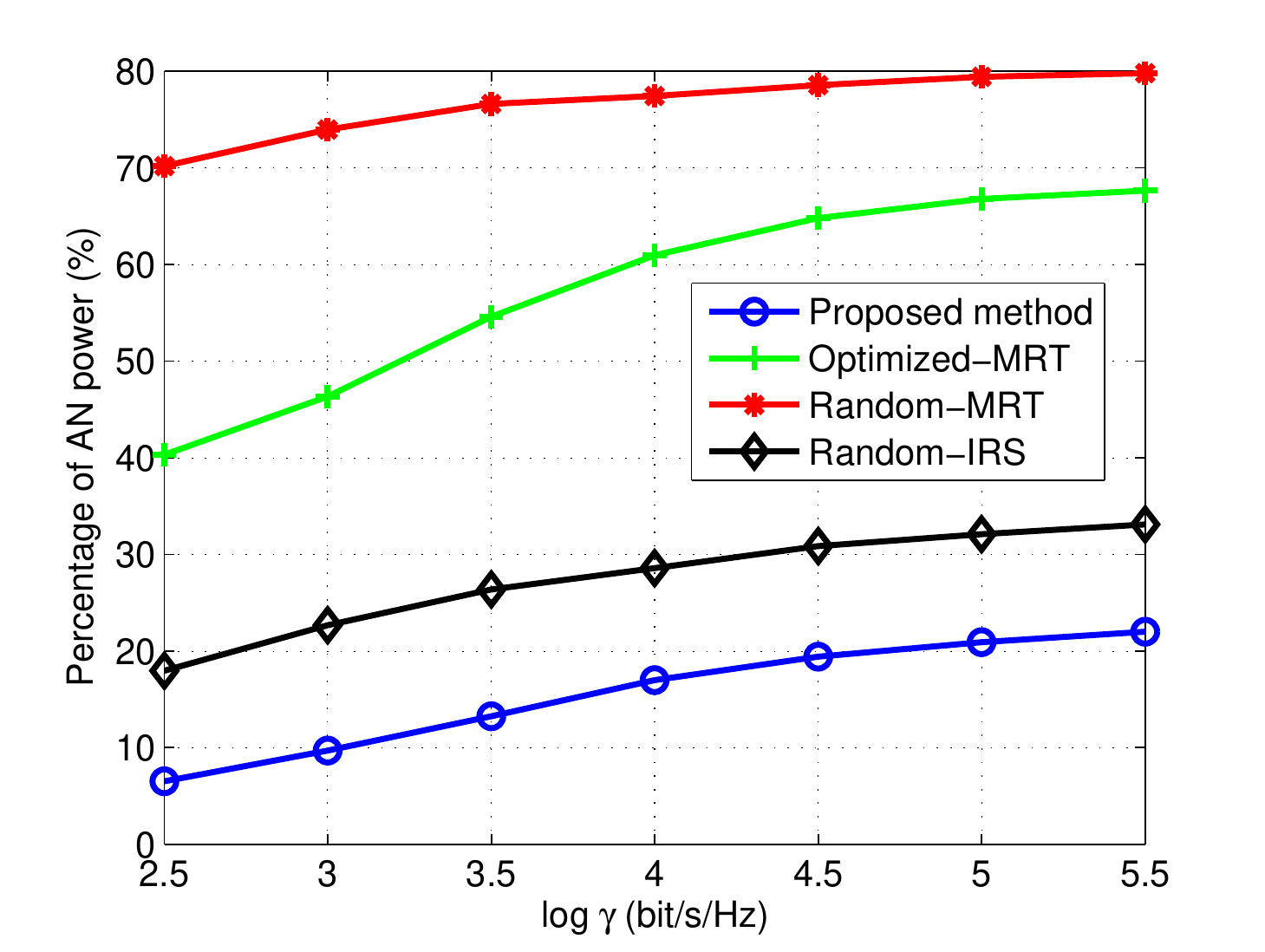}\vspace{-0.2cm}
\caption{Percentage of AN power versus the minimum channel capacity $\log\gamma$
of the legitimate user with $N_{t}=M=6$, $K=2$, and $\delta_{g,k}^{2}=0.0001$,
$\forall k$. The location of AP is $\textrm{(0, 10) m}$,
the location of IRS is $\textrm{(50, 25) m}$, the location of
Bob is $\textrm{(180, 0) m}$, and the locations of Eves are $\textrm{(20, 0) m}$
and $\textrm{(10, 0) m}$.}\vspace{-0.5cm}
\label{figPercentANvsBobGamma}
\end{figure}

The percentage of AN power in the total transmit power versus the
minimum channel capacity $\log\gamma$ of the legitimate user is shown in
Fig. \ref{figPercentANvsBobGamma}. When $\log\gamma$ is small, a
small portion of transmit power is allocated to transmit AN to
deteriorate the achievable rates of Eves. When $\log\gamma$ increases,
more transmit power should be emitted, thus the data rates of both
Bob and Eve increase. To control the data rates of Eves under the
threshold $\log\beta$, more AN power should be emitted to interfere
the Eves. By comparing the proposed
method with the other benchmark schemes, we find that the Optimized-MRT
and Random-MRT schemes have much higher AN power percentage than the proposed method and the Random-IRS scheme. By optimizing the IRS phase shifts, the AN power percentage of the proposed method further reduces compared with the Random-IRS scheme.

\subsection{Transmit Power Versus the Maximum Tolerable Channel Capacity of the
Eavesdroppers}

\begin{figure}[h!]
\centering
\includegraphics[width=3.5in]{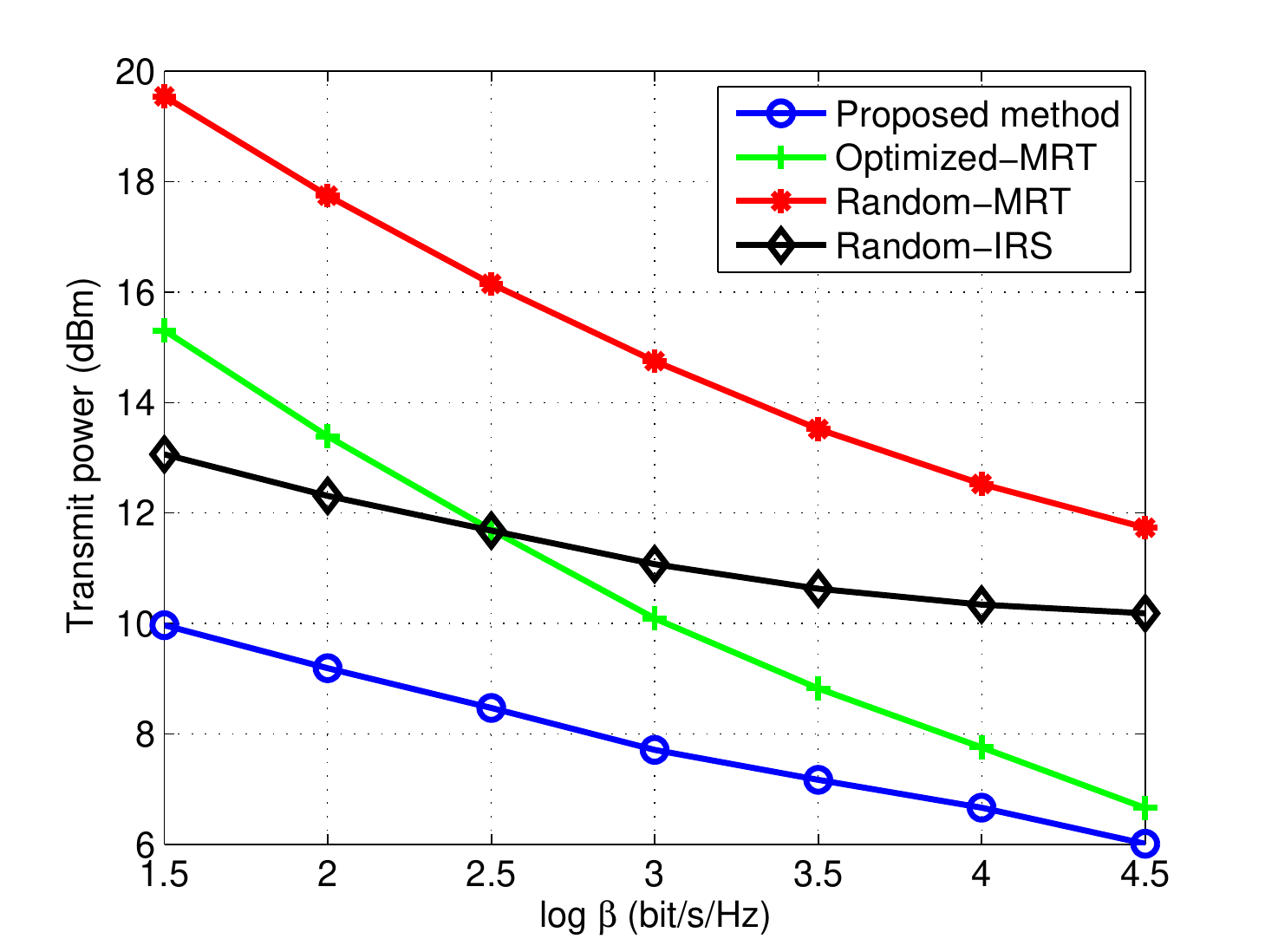}\vspace{-0.2cm}
\caption{Transmit power versus the maximum tolerable channel capacity $\log\beta$
of the eavesdropper with $N_{t}=M=6$, $K=2$ and $\delta_{g,k}^{2}=0.0001$, $\forall k$. The location
of AP is $\textrm{(0, 10) m}$, the location of IRS is $\textrm{(50, 25) m}$,
the location of Bob is $\textrm{(180, 0) m}$, and the locations
of Eves are $\textrm{(20, 0) m}$ and $\textrm{(10, 0) m}$.}\vspace{-0.5cm}
\label{figPvsEveBeita}
\end{figure}

Fig. \ref{figPvsEveBeita} depicts the transmit power versus the maximum tolerable channel
capacity $\log\beta$ of the Eves by assuming that the minimum channel capacity
of the legitimate user is $\log\gamma$=4.7 bit/s/Hz. As shown in Fig. \ref{figPvsEveBeita}, when $\log\beta$
becomes large, more information leakage is allowed for the Eves, and
less AN noise power is required. Thus the transmit power decreases
versus the increased maximum tolerable channel capacity $\log\beta$ of Eves. When $\log\beta$ becomes larger, the consumed transmit power of the Random-MRT scheme approaches that of the Random-IRS scheme, while the consumed transmit power of the Optimized-MRT scheme approaches that of the proposed algorithm. This is because, when more information leakage to Eves is allowed, the importance of robust transmission design by exploiting the Eves' CSI is weakened. By comparing with other benchmark schemes, the proposed method consumes the lowest transmit power, which confirms the effectiveness of the proposed algorithm.

\begin{figure}[h!]
\centering
\includegraphics[width=3.5in]{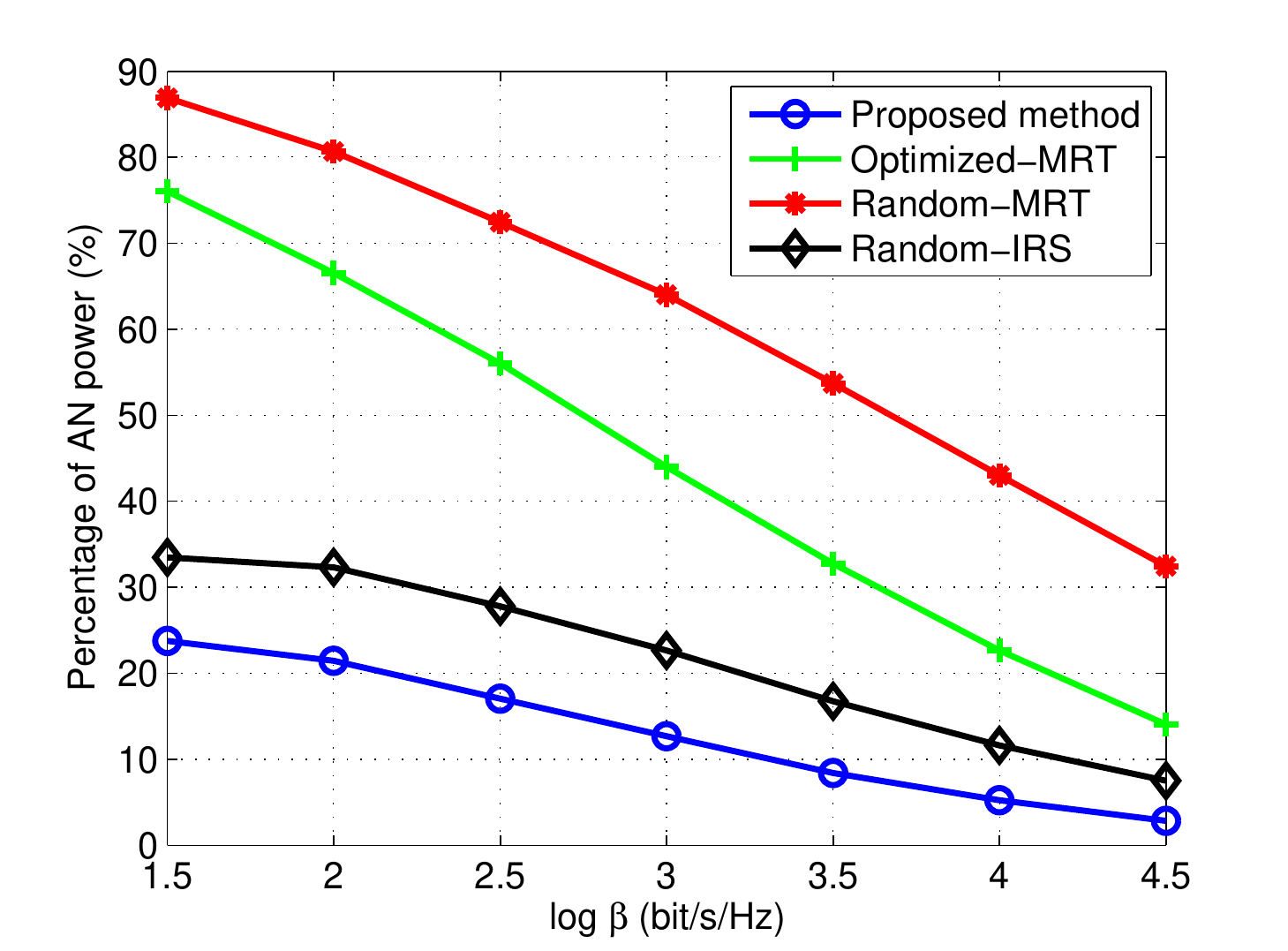}\vspace{-0.2cm}
\caption{Percentage of AN power versus the maximum channel capacity $\log\beta$
of the eavesdropper with $N_{t}=M=6$, $K=2$ and $\delta_{g,k}^{2}=0.0001$, $\forall k$. The location
of AP is $\textrm{(0, 10) m}$, the location of IRS is $\textrm{(50, 25) m}$,
the location of Bob is $\textrm{(180, 0) m}$, and the locations
of Eves are $\textrm{(20, 0) m}$ and $\textrm{(10, 0) m}$. }\vspace{-0.5cm}
\label{figPercentANvsEveBeita}
\end{figure}
The percentage of AN power in the total transmit power versus the maximum tolerable channel
capacity $\log\beta$ of the Eves is depicted
in Fig. \ref{figPercentANvsEveBeita}. When $\log\beta$ becomes large, the percentages of AN power for all methods reduce. It is shown that when $\log\beta$ is small, a large portion
of the transmit power is allocated to transmit AN to deteriorate
the achievable rates of the potential eavesdroppers, and therefore,
there is less power transmitted for legitimate user. As $\log\beta$
increases, the constraints on the performance of the Eves are relaxed,
and more transmit power is allocated to beamform to Bob. The percentages of AN power of both Random-MRT and Optimized MRT schemes are higher than those of the Random-IRS and the proposed method. This is because the channel information of Eves is not utilized for the MRT schemes. The percentage of AN power for the proposed method is the lowest, which demonstrates the effectiveness of the robust transmission design.

\subsection{Transmit Power Versus the Number of IRS Elements}

\begin{figure}[h!]
\centering
\includegraphics[width=3.5in]{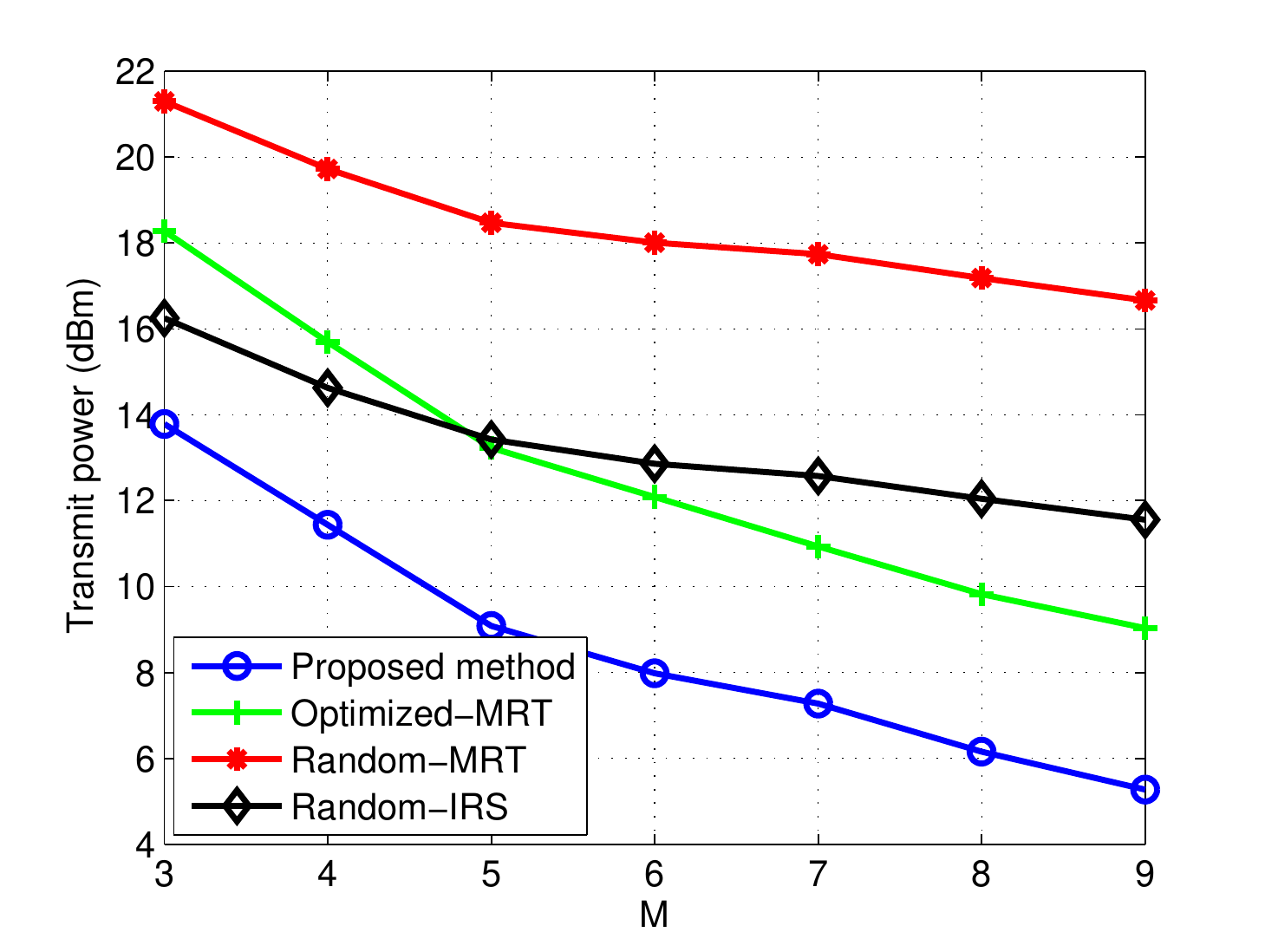}\vspace{-0.2cm}
\caption{Transmit power versus the IRS element number $M$ with $N_{t}=3$, $K=2$, $\delta_{g,k}^{2}=0.0001$, $\forall k$, $\log\beta=1$ $\textrm{bit/s/Hz}$ and $\log\gamma=3$ $\textrm{bit/s/Hz}$. The location of AP is $\textrm{(0, 10) m}$, the location of IRS is $\textrm{(100, 25) m}$, the location of
Bob is $\textrm{(180, 0) m}$, and the locations of Eves are $\textrm{(160, 0) m}$
and $\textrm{(170, 0) m}$.}\vspace{-0.5cm}
\label{figPvsM}
\end{figure}

We further examine the minimum transmit power consumption versus different numbers of IRS elements in Fig. \ref{figPvsM}. The maximum
tolerable channel capacity of the Eves is $\log\beta=1$ $\textrm{bit/s/Hz}$.
The required minimum data rate for the Bob is set as $\log\gamma=3$
bit/s/Hz. The transmit power is reduced when the IRS element number
$M$ increases. To enhance the security, the phase shifts of the IRS can be optimized to help the data transmission for Bob while degrade the data transmission for Eves. A large value of $M$ will bring more degrees of freedom to adjust the cascaded channels, thus less transmit power is required to guarantee the data transmission for Bob and impair the data transmission for Eves.
The transmit power of the Optimized-MRT scheme and the proposed algorithm decreases more quickly than that of the Random-MRT scheme and Random-IRS scheme, which shows that optimizing the phase shifts of IRS can reduce the transmit power effectively. The transmit power of the Random-MRT and Random-IRS schemes decreases slowly when $M$ is large, which shows that the transmit power cannot be effectively reduced without IRS phase shift optimization. Compared with the other benchmark schemes, the proposed algorithm requires the lowest transmit power, which validates the superiority of the proposed algorithm.

\subsection{Transmit Power Versus the Number of Transmit Antennas}

\begin{figure}[h!]
\centering
\includegraphics[width=3.5in]{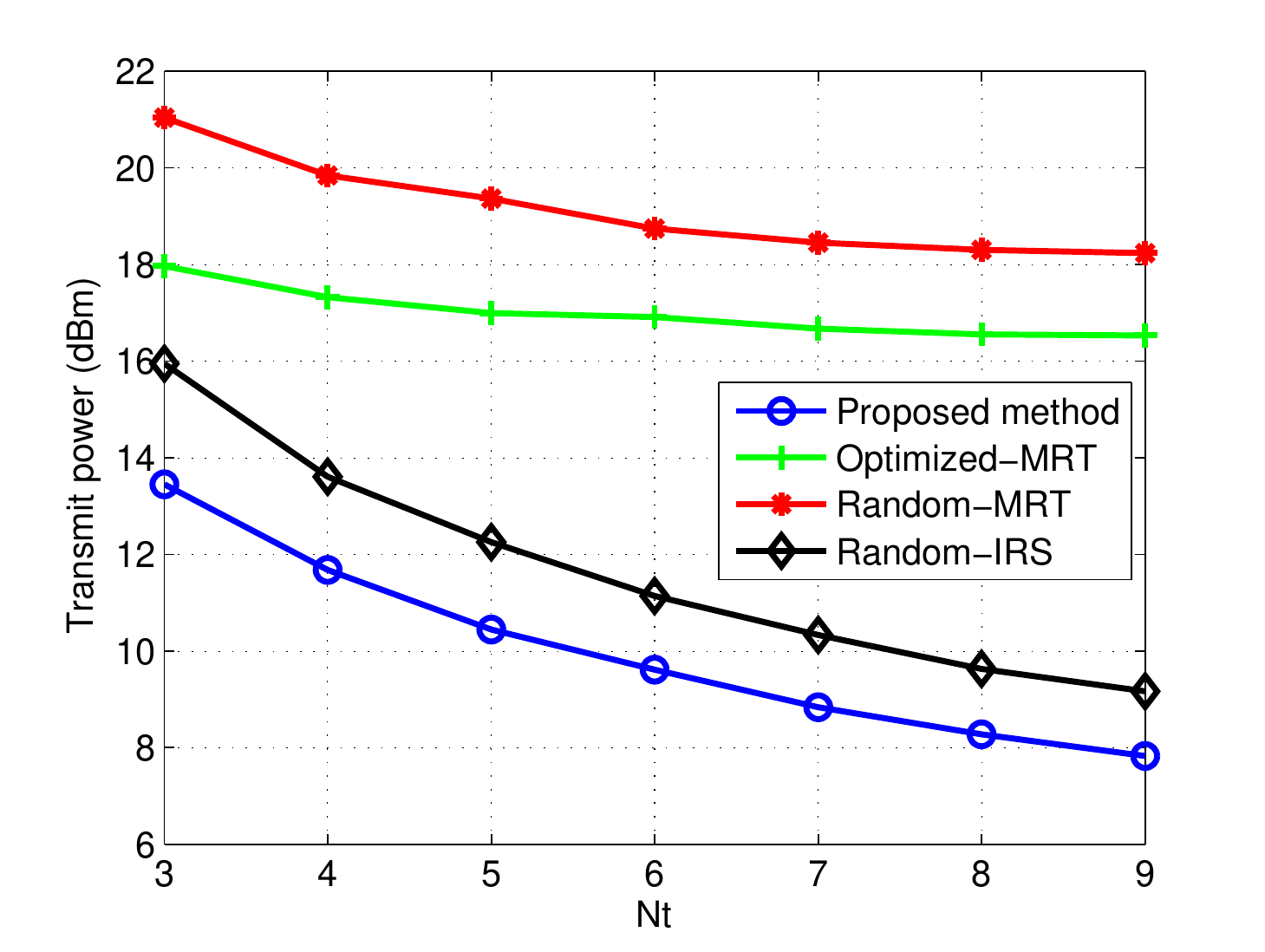}\vspace{-0.2cm}
\caption{Transmit power versus transmit antenna number $N_{t}$ with $M=3$, $K=2$, $\delta_{g,k}^{2}=0.0001$, $\forall k$, $\log\beta=1$ $\textrm{bit/s/Hz}$ and $\log\gamma=3$ $\textrm{bit/s/Hz}$. The location of AP is $\textrm{(0, 10) m}$,
the location of IRS is $\textrm{(100, 25) m}$, the location of
Bob is $\textrm{(180, 0) m}$, and the locations of Eves are $\textrm{(160, 0) m}$
and $\textrm{(170, 0) m}$. }\vspace{-0.5cm}
\label{figPvsNt}
\end{figure}

We further examine the transmit power consumption versus different number of transmit antennas $N_{t}$ in Fig. \ref{figPvsNt}. The maximum tolerable
channel capacity of the Eves is $\log\beta=1$ $\textrm{bit/s/Hz}$. The
required minimum data rate for the Bob is set as $\log\gamma=3$
bit/s/Hz. Fig. \ref{figPvsNt} shows that the transmit power of all schemes reduces when the number of transmit antennas increases. A large value of $N_t$ helps improve the channels for both the Bob and Eves. Then, the robust beamforming design for Bob and AN design for Eves are easier to achieve with a larger $N_t$. Thus, the required transmit power reduces with increased $N_t$. Compared with the Random-MRT scheme and Optimized-MRT scheme, the required transmit power of the Random-IRS scheme and proposed algorithm decreases more quickly with $N_t$. This is because, the MRT schemes do not exploit the Eves' CSI, while the Random-IRS scheme and the proposed algorithm exploit the Eve's CSI robustly. The required transmit power of Random-IRS scheme is further reduced by the proposed algorithm, which validate the effectiveness of the proposed robust transmission design.

\subsection{Transmit Power Versus CSI Uncertainty}

\begin{figure}[h!]
\centering
\includegraphics[width=3.5in]{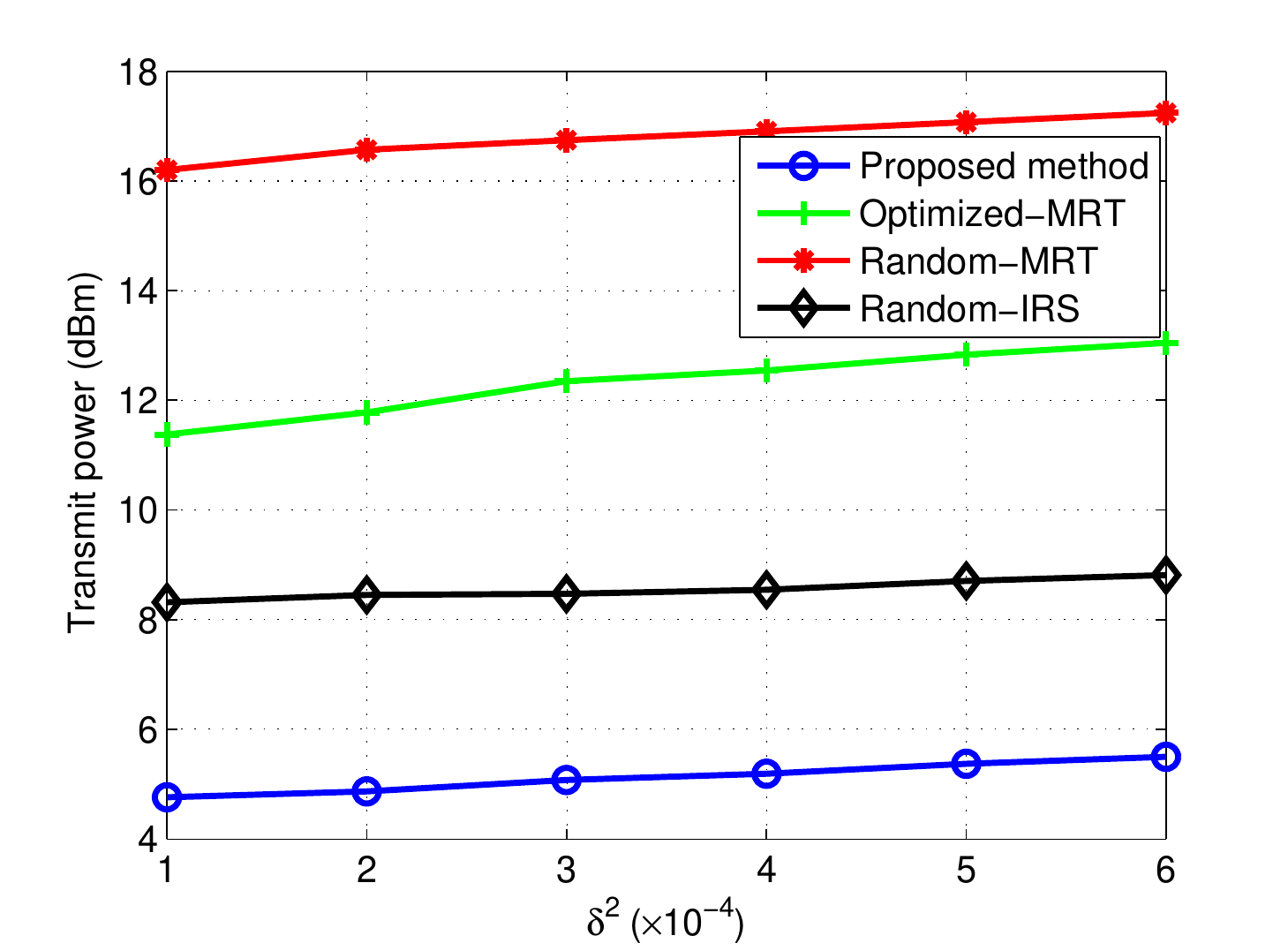}\vspace{-0.2cm}
\caption{Transmit power versus normalized CSI error variance $\delta^{2}$ with $N_t=M=6$, $K=2$, $\log\beta=1$ $\textrm{bit/s/Hz}$ and $\log\gamma=3$ $\textrm{bit/s/Hz}$. The location of AP is $\textrm{(0, 10) m}$,
the location of IRS is $\textrm{(100, 25) m}$, the location of
Bob is $\textrm{(180, 0) m}$, and the locations of Eves are $\textrm{(160, 0) m}$
and $\textrm{(170, 0) m}$.}\vspace{-0.5cm}
\label{figPvsCSIerr}
\end{figure}

Fig. \ref{figPvsCSIerr} shows the transmit power versus the normalized
CSI error variance $\delta^{2}$, where the normalized CSI errors of different eavesdroppers are assumed to be identical, i.e., $\delta_{g,k}=\delta$, $\forall k$. As can be observed, for the proposed
scheme and the other benchmark schemes, the transmit power increases
as the quality of the CSI degrades. When the CSI errors of Eves' channels increase, more AN power is required to interfere Eves for information leakage limitation, thus the allocated power for beamforming must also increase to guarantee Bob's data requirement. This leads to the increase of total transmit power with deteriorating channel. In comparison to other three benchmark schemes, the required transmit power of the proposed method has been significantly reduced. Both the Random MRT and Optimized MRT schemes need much more transmit power to impair the Eves, which shows that the utilization of Eves' CSI can help reduce the transmit power. The Random IRS scheme also needs much more transmit power than the proposed method, which indicates the effectiveness of the optimization on
the IRS phase shifts in the robust transmission design. It can also be found from Fig. \ref{figPvsCSIerr} that the transmit power of the Random MRT and Optimized MRT schemes increases more quickly than the proposed method, which shows that they are more sensitive to the CSI errors, and that the proposed method is more robust with respect to the CSI error.

\begin{figure}[h!]
\centering
\includegraphics[width=3.5in]{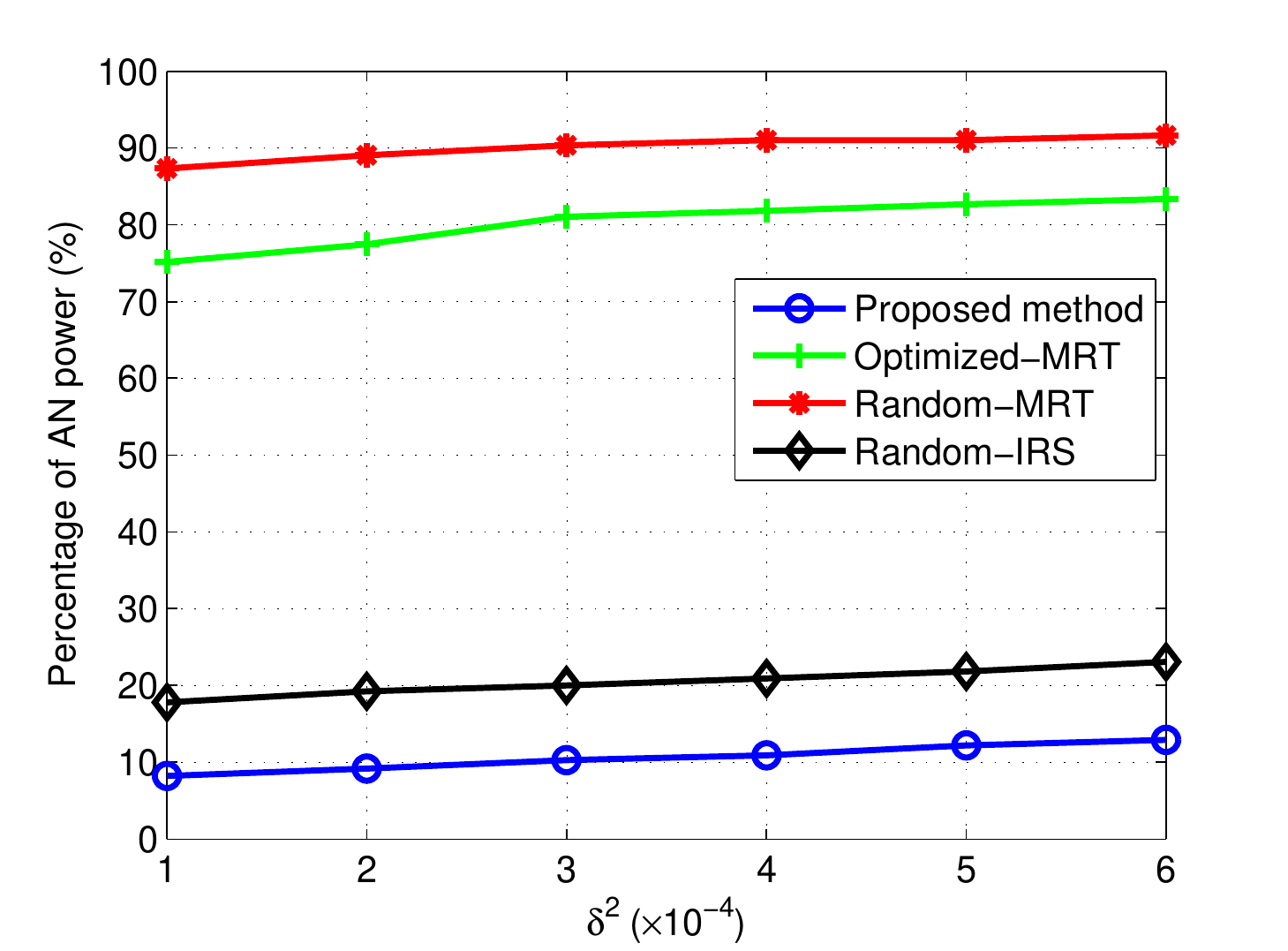}\vspace{-0.2cm}
\caption{Percentage of AN power versus normalized CSI error variance $\delta^{2}$ with $N_t=M=6$, $K=2$, $\log\beta=1$ $\textrm{bit/s/Hz}$ and $\log\gamma=3$ $\textrm{bit/s/Hz}$. The location of AP is $\textrm{(0, 10) m}$,
the location of IRS is $\textrm{(100, 25) m}$, the location of
Bob is $\textrm{(180, 0) m}$, and the locations of Eves are $\textrm{(160, 0) m}$
and $\textrm{(170, 0) m}$.}\vspace{-0.5cm}
\label{figPercentANvsCSIerr}
\end{figure}

Fig. \ref{figPercentANvsCSIerr} gives the percentage of AN power
in the total transmit power versus the normalized CSI error variance
$\delta^{2}$. It is shown in Fig. \ref{figPercentANvsCSIerr} that
the percentage of AN power increases with CSI estimation error.
When the quality of the CSI degrades, more AN power is required to
interfere the eavesdroppers. The percentages of AN power in the total transmit power for the Optimized MRT and Random MRT schemes are more than 70$\%$, while the percentages of AN power in the total transmit power for the proposed method and Random IRS scheme are less than 20$\%$. Without the Eves's CSI, more AN power is required to impair the Eves. On the contrary, by exploiting the Eves' CSI robustly, the AN interference becomes more effective, thus less percentage of AN power is required. Compared with the Random IRS scheme, less AN power percentage is required for the proposed method, which demonstrates that the optimization of IRS phase shifts helps impair the Eves' channel, thus makes less AN power interferes the Eves more effectively.

\section{Conclusion}

In this paper, we designed the outage constrained robust transmission
in the secure IRS-aided wireless communications. The statistical CSI error of cascaded IRS channel was taken
into consideration for the first time in secure communications, and
the OC-PM problem was formulated to minimize the transmit power by jointly optimizing the transmit beamforming vector, AN covariance, and IRS
phase shifts. To solve it, an AO based method was proposed,
where the optimization variables are optimized alternately. The BTI was utilized to tackle the chance constraint. The SDR scheme was utilized when designing the
beamforming vector and IRS phase shifts. Specifically, the nonconvex
unit modulus constrained was handled by a penalty method. Simulation
results have verified the effectiveness of IRS on enhancing the physical
layer security of wireless communications. The robustness of the proposed
method has also been confirmed when the statistical CSI error exists, which
is meaningful. The superiority of the proposed method
over the benchmark methods has also been validated.

\vspace{-0.5cm}
%\vspace{-1.2cm}
\bibliographystyle{IEEEtran}
\bibliography{OC-PMRef}

%\bibliographystyle{plain}
%\addcontentsline{toc}{section}{\refname}\bibliography{Myref_OC_SRM}

\end{document}